\documentclass[
 reprint,
superscriptaddress,
 amsmath,amssymb,
 aps,
pra,
nofootinbib, longbibliography]{revtex4-1}
\usepackage{hyperref}

\usepackage{graphicx}
\usepackage{dcolumn}
\usepackage{bm}
\usepackage{verbatim}
\usepackage{float}
\usepackage{subcaption}

\usepackage{balance}
\usepackage{amsmath}
\usepackage{amssymb}

\usepackage{color}

\begin{document}

\title{Polaritonic ultracold reactions: cavity controlled molecular photoassociation} 

\author{Vasil~Rokaj}
\email{vasil.rokaj@cfa.harvard.edu}
\affiliation{ITAMP, Center for Astrophysics $|$ Harvard $\&$ Smithsonian, Cambridge, USA}
\affiliation{Department of Physics, Harvard University, Cambridge, USA}

\author{Simeon~I.~Mistakidis}
\affiliation{ITAMP, Center for Astrophysics $|$ Harvard $\&$ Smithsonian, Cambridge, USA}
\affiliation{Department of Physics, Harvard University, Cambridge, USA}

\author{H.~R.~Sadeghpour}
\affiliation{ITAMP, Center for Astrophysics $|$ Harvard $\&$ Smithsonian, Cambridge, USA}

\date{\today}

\begin{abstract}
We introduce a prototypical model for cavity polaritonic control of ultracold photochemistry by considering the resonant vibrational strong coupling of a rubidium dimer ($\rm{Rb_2}$) to a terahertz cavity. We demonstrate that at avoided crossings between a vibrational excitation and the vacuum photon absorption, the resulting polaritonic states between the molecule and photons can efficiently control the molecular vibrational Franck-Condon (FC) factors. Due to the entanglement between light and matter, FC factor is transferred from one polaritonic branch to the other, leading to a polariton with a substantially enhanced FC factor. Utilizing this polariton state for photoassociation results in the enhanced formation of ultracold molecules. This work suggests a path to controlling photoassociation with cavity vacuum fields, and lays the ground for the emerging subfield of polaritonic ultracold chemistry.
 \end{abstract}

\pacs{Valid PACS appear here}
\maketitle

\begin{figure}[H]
    \centering
    \includegraphics[width=\columnwidth]{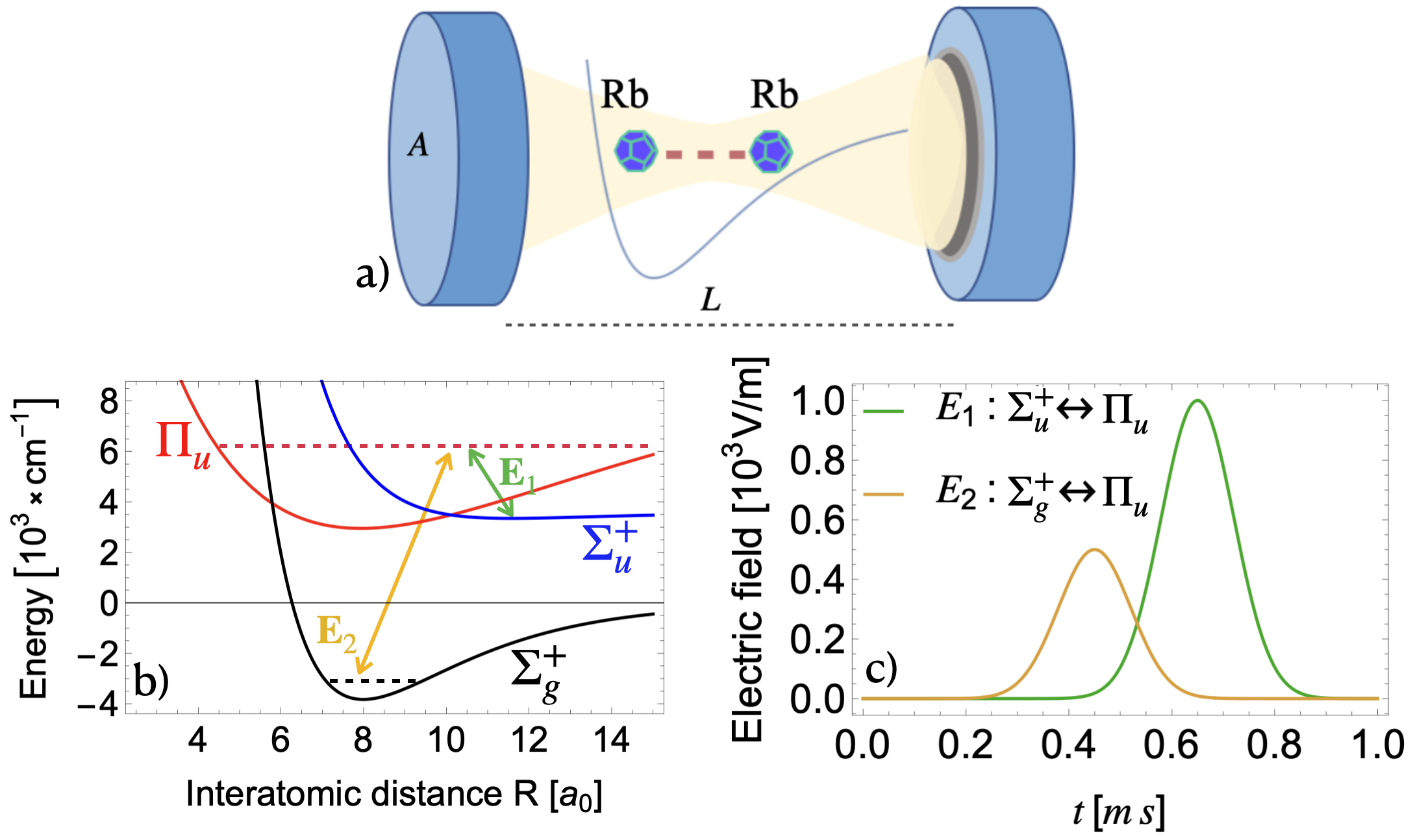}
    \caption{a) Schematic depiction of the $\textrm{Rb}_2$ molecule coupled to a cavity mode. The distance between the cavity mirrors is $L$ and the area of the mirrors $A$. b) Potential energy surfaces of $\textrm{Rb}_2$ for the $\Sigma^+_u, \Pi_u$ and $\Sigma^+_g$ states. c) Gaussian pulses applied from the open sides of the cavity for the photoassociation of $\textrm{Rb}_2$ molecules to the vibrational ground state in $\Sigma^+_g$.}
    \label{Setup}
\end{figure}

Achieving two of the holy grails of chemistry \cite{wilson95}, manipulation of matter at the atomic level \cite{eigler90,sugimoto05,gross09}, and coherent control over chemical pathways \cite{tannor86,shapiro03,rabitz88} has made great strides in the last few decades. Meanwhile, quantum leap advances in cooling and trapping of atoms and molecules are bringing these dreams closer to reality, as single atoms and molecules can be now trapped in periodic arrays, created by optical lattices or tweezers~\cite{Doyletweezer, tweezermolecules, Anderegg2018, Park2023}. These advances have enabled the application of ultracold molecules in quantum chemistry~\cite{NiColdCollisions}, searches for fundamental physics~\cite{ClarkRMP}, quantum simulation~\cite{DeMilleQComp,MoleculeSimulation, Doyletweezer}, and sensing~\cite{YeReview, Julieneundercontrol}. This wide range of applications renders the production of ultracold molecules highly desirable. Typically, for the enhancement of molecular photoassociation the stimulated Raman adiabatic passage (STIRAP) is employed, which consists of a so-called counterintuitive succession of two laser pulses driving molecular transitions~\cite{DenschlagPRL, StirapRMP} as it is illustrated in Fig.~\ref{Setup}(c).   

Recently, the possibility to control chemical reactions~\cite{hutchison2012, ebbesen2016, hutchison2013,  orgiu2015, galego2016, flick2017, schafer2019modification} and material properties~\cite{Bloch2022Review, SchlawinSentefReview, paravacini2019, FaistCavityHall, Rokaj2deg, rokaj2023topological} with cavity vacuum fields has emerged. The experimental demonstration of vibrational strong coupling and resonant modification of reaction rates~\cite{hutchison2012, BlakeScience} indicates that chemistry can be affected by the quantum hybridization of molecular states and cavity photons, known as polaritons~\cite{RuggiReview2023, JoelReview, HuoReview2023, VidalEbbesenReview}. The polaritons formed in vacuum cavities are free from external driving fields that can lead to thermalization or destroy quantum coherence~\cite{BarnesReview2022}. 

It may now be possible to combine some of the unique elements of ultracold molecular physics with cavity quantum electrodynamics (cQED) to steer photochemistry through vibrational polariton states. In this Letter, we present a prototypical model of an ultracold reaction under strong coupling to the cavity field. We showcase that vibrational polariton states, formed between the $\textrm{Rb}_2$ dimer and the cavity photons, can be utilized as a new paradigm for the controlled photoassociation of molecules. 
We find that at the avoided crossings, the transition probabilities between the vibrational states, i.e., the FC factors, of the polaritons are significantly modified. More precisely, in the vicinity of an avoided crossing, one polariton branch acquires an enhanced FC factor while the FC factor for the other polariton branch is suppressed. We name this phenomenon polariton FC transfer, as the population is transferred from one polariton state to the other. We show that the polariton FC transfer is a result of entanglement between the photon and molecule excitations. The FC factors determine the molecular yield. By utilizing the polariton branch with the enhanced FC factors for STIRAP, it is shown that molecular photoassociation of $\textrm{Rb}_2$ is substantially enhanced when the molecules are coupled to the cavity field.

\textit{Rubidium Dimer in the Cavity.}---We consider the $\textrm{Rb}_2$ coupled to a single-mode terahertz cavity~\cite{hutchison2012, ebbesen2016, hutchison2013, BlakeScience}. The molecule-photon interacting system is described by the length gauge Hamiltonian~\cite{rokaj2017, cohen1997photons}
\begin{eqnarray}\label{Hamiltonian}
\mathcal{H}=\mathcal{H}_M+\hbar\omega a^{\dagger}a-\sqrt{\frac{\hbar \omega}{2}}\lambda d_R\left(a+a^{\dagger}\right)  +\frac{\lambda^2 d^2_R}{2}.
\end{eqnarray}
Here, $\mathcal{H}_{M}= -\frac{\hbar^2 }{2\mu}\frac{\partial^2}{\partial R^2}+D\left[e^{-2w(R-R_0)}-2e^{-w(R-R_0)}\right]$ represents the molecular Hamiltonian of $\textrm{Rb}_2$. All potential energy surfaces (PESs) of $\textrm{Rb}_2$ are described with the Morse potential~\cite{Morse} characterized by the experimental parameters reported in Ref.~\cite{RB2potentials}, thus rendering our molecular model quantitatively precise. The reduced mass of the dimer is $\mu$, $D$ ($w$) denotes the depth (width) of the potential well, and $R_0$ is the equilibrium bond distance. Further, we assume a linear dipole operator $d_R=\mu_0R$ with $\mu_0$ being the strength of the molecular dipole moment. The length gauge Hamiltonian contains both the bilinear coupling, and the quadratic dipole self-energy, $\sim \lambda^2 d^2_R$, which is necessary for the stability of the molecular system and for preserving gauge invariance~\cite{rokaj2017, bernadrdis2018breakdown}. We note that in Eq.~$\ref{Hamiltonian}$ the trivial zero-point energy of the cavity mode has been removed. The cavity field strength $\lambda=\sqrt{1/\epsilon_0 \mathcal{V}}$ depends on the cavity volume $\mathcal{V}=AL=A\pi c/\omega$ with $A$ being the area of the cavity mirrors having a distance $L$ and $\omega=\pi c/L$ is the frequency of the cavity mode. Further, $c$ is the speed of light, $\epsilon_0$ the vacuum permittivity, and $a, a^{\dagger}$ are bosonic operators of the cavity mode satisfying $[a,a^{\dagger}]=1$. The independent cavity-related parameters are the frequency $\omega$ and the light-matter coupling strength $g=\mu_0\lambda/\sqrt{\omega}=\mu_0/\sqrt{A\pi c\epsilon_0}$ (see the Supplementary Material (SM) for more details~\cite{supp}).

\begin{figure}[H]
    \centering
    \includegraphics[width=\linewidth]{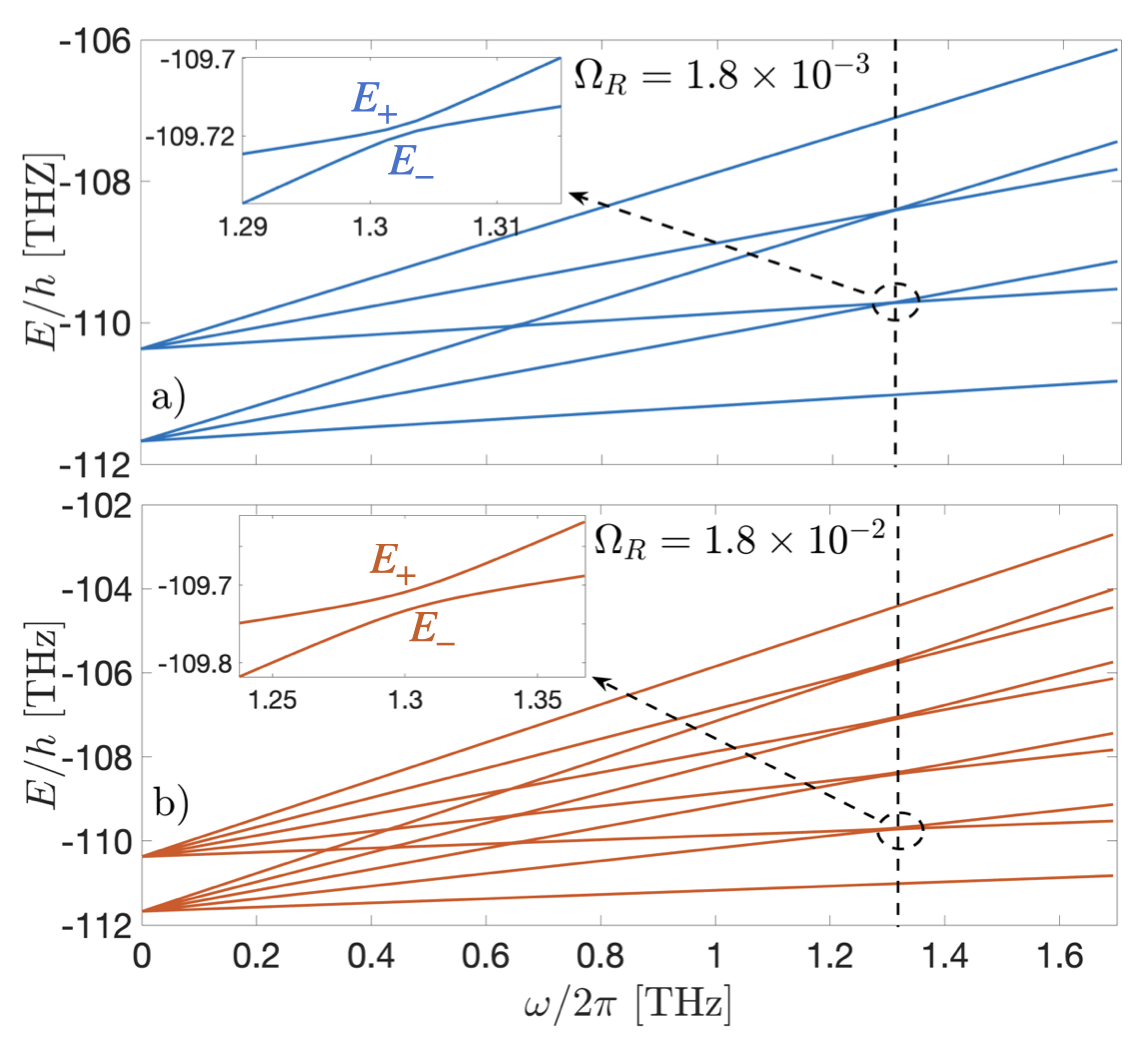}
    \caption{Polaritonic energy spectra for the vibrational states $n=60$ and $m=61$ in the $\Pi_u$ PES coupled to a cavity mode. In a) the light-matter coupling strength is $g=0.001 \textrm{a.u.}$ and three photonic Fock states are needed to converge the first avoided crossing (inset) which exhibits normalized Rabi splitting $\Omega_R=1.8\times 10^{-3}$. In b) $g=0.01 \textrm{a.u.}$ is enhanced and 5 photonic Fock states are required to converge the first avoided crossing (inset) which exhibits an order of magnitude larger normalized Rabi splitting $\Omega_R=1.8\times 10^{-2}$.}
    \label{Spectra}
\end{figure}

The $\textrm{Rb}_2$ molecule is homonuclear, and thus in the $\Sigma^+_{g}$ and the $\Sigma^+_{u}$ PESs it has no permanent dipole. Consequently, the cavity field does not couple the vibrational states in these PESs, but rather couples only with the $\Pi_u$ PES where $\textrm{Rb}_2$ has a dipole moment. To treat the photon-molecule system we expand the Hamiltonian in the basis consisting of the vibrational eigenstates of the $\Pi_u$ PES $\{|\Phi^{\Pi_u}_n\rangle\}$ and the eigenstates $\{|i\rangle\}$ of the bosonic cavity mode~\cite{GriffithsQM}. As such, we obtain the matrix form $\mathcal{H}^{ji}_{mn}:=\langle j|\langle \Phi^{\Pi_u}_m|\mathcal{H}|\Phi^{\Pi_u}_n\rangle|i\rangle$ of the Hamiltonian $\mathcal{H}$. Diagonalizing numerically $\mathcal{H}^{ji}_{mn}$ we find the vibrational polaritonic energy levels and the hybrid polaritonic states. For convenience, the Hamiltonian $\mathcal{H}$ is set up in atomic units (a.u.), but then all observables are provided in units of Hertz. We restrict ourselves to weak and strong light-matter coupling strengths quantified by  $g=10^{-3}\textrm{a.u.}$ and $g=10^{-2}\textrm{a.u.}$ respectively. The corresponding polariton energy spectra, for these $g$ values, are shown in Fig.\ref{Spectra} as a function of the cavity frequency for the vibrational levels $n=60$ and $m=61$ in the $\Pi_u$ PES.
For both coupling strengths at the point of resonance indicated by the vertical dashed line in Figs.\ref{Spectra}(a) and (b), an avoided crossing emerges which manifests the hybridization between light and matter. For weak (strong) coupling, the lowest energetically avoided crossing presented in the inset of Fig.~\ref{Spectra}(a) (Fig.~\ref{Spectra}(b)) features the normalized Rabi splitting $\Omega_R=(E_+-E_-)/\hbar \omega_{\rm res}=0.0018$ ($\Omega_R=0.018$). This enhanced normalized Rabi splitting,  $\Omega_R=0.018$, indicates the entrance to the strong coupling regime, at which the frequency window where the avoided crossing occurs becomes larger.
This implies that light-matter entanglement persists for a wider region around the resonance point as can be readily seen by comparing the insets in Figs.~\ref{Spectra}(a) and (b).

\begin{figure}[H]
    \centering
    \includegraphics[width=0.9\columnwidth]{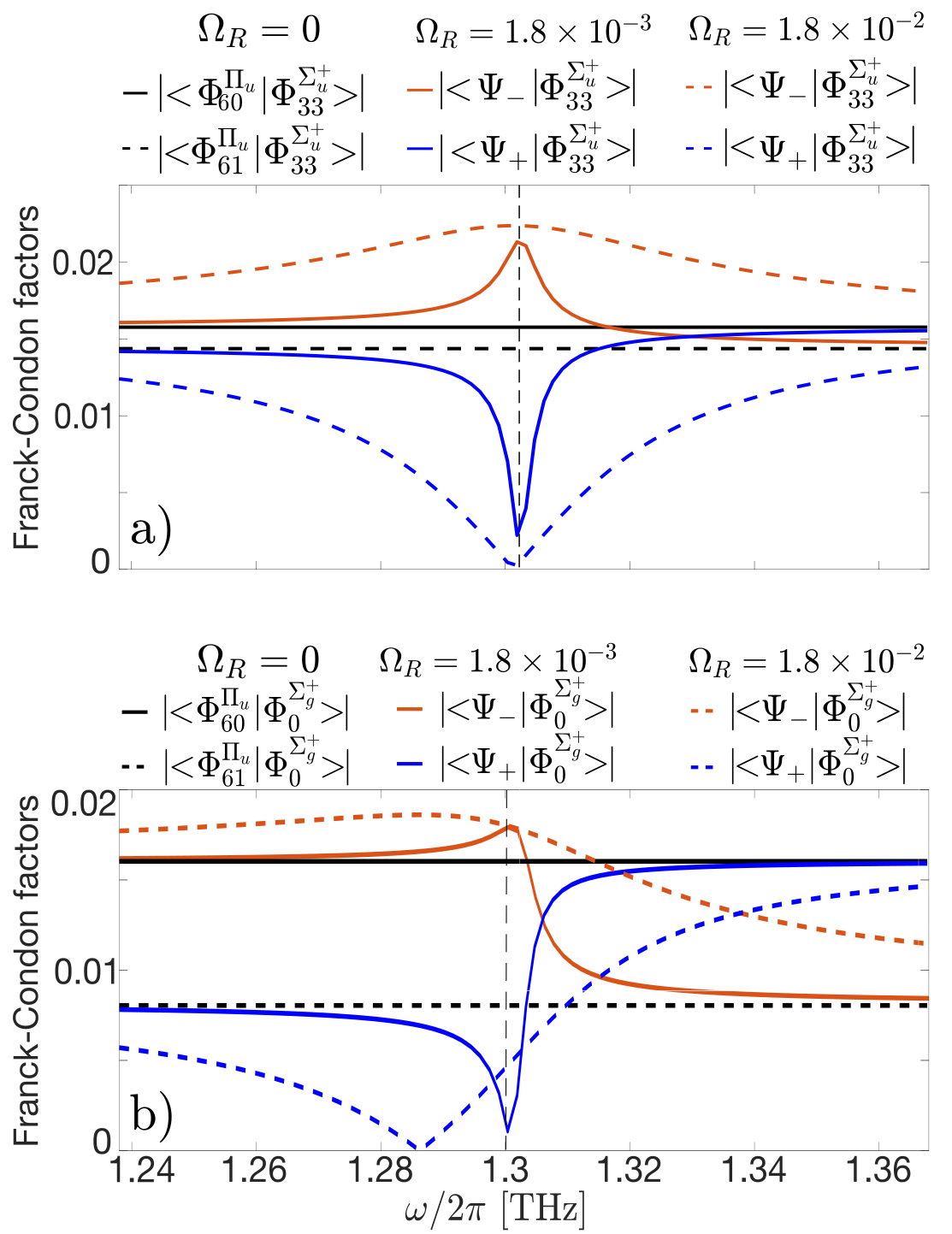}
    \caption{Bare and polaritonic FC factors of $\rm{Rb}_2$ for normalized Rabi splitting $\Omega_R=1.8\times 10^{-3}$ and $\Omega_R=1.8\times 10^{2}$. In a) the FC factors for the upward transition from $\Phi^{\Sigma^+_u}_{33}$ to $\Phi^{\Pi_u}_{60,61}$ and the respective polariton states $|\Psi_{\pm}\rangle$ in $\Pi_u$ are shown (see legend). In b) the FC factors for the downward transition to the ground state $\Phi^{\Sigma^+_g}_0$, starting from $\Phi^{\Pi_u}_{60,61}$ and the polariton states $|\Psi_{\pm}\rangle$ are presented (see legend). As it can be seen, at resonance the lower (upper) polariton $\Psi_-$ ($\Psi_+$) shows enhanced (suppressed) FC factors. Importantly, for strong coupling, $\Omega_R=1.8\times 10^{-2}$, polariton FC transfer is enhanced.}
    \label{PFCs}
\end{figure}

\textit{Polariton Franck-Condon Transfer.}---The FC factors are decisive for the photoassociation of ultracold molecules, as they determine the transition probabilities between vibrational states. The polariton states in the $\Pi_u$ PES have the form $|\Psi_{p}\rangle=\sum _{n,i}c^p_{ni}|\Phi^{\Pi_u}_n\rangle|i\rangle$ with the coefficients $c^p_{ni}$ obtained from the diagonalization of $\mathcal{H}^{ji}_{mn}$. To compute the FC factors between the vibrational polariton states $|\Psi_{p}\rangle$ in the $\Pi_u$ PES, and the vibrational states in the $\Sigma^+_g$ and $\Sigma^+_u$ PESs  we need to construct the tensor product state between the vibrational states $|\Phi^{\Sigma^+_{g,u}}_{v}\rangle$ and the photonic unit vector $|\mathbb{I}\rangle=\sum^{l}_{i=0}|i\rangle$.
Here, $l$ denotes the amount of photonic Fock states included in the computation of the vibrational polaritons. Then, the overlap $ \langle \Phi^{\Sigma^+_{g,u}}_{v}|\Psi_p\rangle :=\langle\mathbb{I}|\langle \Phi^{\Sigma^+_{g,u}}_{v}|\Psi_p\rangle$ provides the polariton FC factors. Fig.~\ref{PFCs} depicts the bare, $\Omega_R=0$, and the respective polaritonic FC factors at the lowest avoided crossing for $\Omega_R=0.0018$ and $\Omega_R=0.018$ with respect to the vibrational levels $n=60$ and $m=61$ in $\Pi_u$. Two particular transitions are shown, from the last vibrational state in $\Sigma^+_u$ PES, $\Phi^{\Sigma^+_u}_{33}$, and to the vibrational ground state $\Phi^{\Sigma^+_g}_0$ as a function of the cavity frequency. These states are of specific interest since they will be deployed later for the STIRAP photoassociation. In both panels of Fig.~\ref{PFCs} we observe that before the resonance point (indicated by the vertical dashed line) the upper (lower) polariton has FC factor equal to the bare vibrational states $m=61$ ($n=60$). 
However,  after the resonance point $\omega/2\pi=1.3\textrm{THz}$ the situation is inverted. The most interesting phenomenon is that at resonance the FC factor of the lower polariton increases substantially while the one of the upper polariton is suppressed. Therefore, through the hybridization and entanglement between the vibrational states with the cavity photons, FC factor is transferred from the upper polariton to the lower one. This phenomenon occurring in the vicinity of the avoided crossing will be named in the following as \textit{polariton Franck-Condon transfer}. It is crucial for the photoassociation of molecules because the FC factors determine the molecular yield. To the best of our knowledge the polariton Franck-Condon transfer has not been reported elsewhere. Importantly, in order to guarrantee that the molecular formation through STIRAP using the polariton states will be enhanced, the FC factors for both transitions, $\Sigma^+_u \leftrightarrow \Pi_u $ and $\Pi_u \leftrightarrow \Sigma^+_g$, of the same polariton branch need to be enhanced.

\textit{Insights from the Jaynes-Cummings model.}---To shed further light on the behavior of the polariton FC factors from an analytical perspective, we employ the Jaynes-Cummings (JC) model. The latter captures the hybridization of two vibrational levels by a photon in the weak coupling regime, $\Omega_R\sim 10^{-3}$,~\cite{shore1993}. However, for strong coupling, $\Omega_R\sim 10^{-2}$, the interactions beyond the rotating-wave approximation become substantial and the JC model is no longer quantitatively precise, see SM for further details~\cite{supp}. The hybrid excited states in the JC model for single photon excitation exactly at resonance are $|\pm\rangle= (|\Phi^{\Pi_u}_{n}\rangle |1\rangle \mp |\Phi^{\Pi_u}_{n+1}\rangle|0\rangle)/\sqrt{2}$~~\cite{scully1997, Vogel}. Their FC factors with the vibrational states $|\Phi^{\Sigma^+_{g,u}}_{v}\rangle$ are defined as $\textrm{FC}_{\pm}=\langle \mathbb{I}| \langle\Phi^{\Sigma^+_{g,u}}_{v} |\pm\rangle$ and can be expressed as 
\begin{equation}\label{DressedFC}
    \textrm{FC}_{\pm}=\frac{1}{\sqrt{2}}\left[\langle \Phi^{\Sigma^+_{g,u}}_{v}|\Phi^{\Pi_u}_{n}\rangle \mp \langle \Phi^{\Sigma^+_{g,u}}_{v}|\Phi^{\Pi_u}_{n+1}\rangle \right].
\end{equation}
It becomes evident that at resonance the FC factors of the JC states can be significantly modified. Namely, if the vibrational overlaps $\langle \Phi^{\Sigma^+_{g,u}}_{v}|\Phi^{\Pi_u}_{n}\rangle, \langle \Phi^{\Sigma^+_{g,u}}_{v}|\Phi^{\Pi_u}_{n+1}\rangle$ have the same sign then $\textrm{FC}_-$ ($\textrm{FC}_+$) gets enhanced (suppressed). However, if the integrals have opposite signs the situation is inverted. This demonstrates that around the resonance point there is a significant transfer of FC factor from one branch of the avoided crossing to the other. The phenomenon of FC factor exchange is a result of entanglement between the photons and the molecule around the resonance point. Importantly, for the STIRAP through the lower (upper) polariton to be enhanced it is necessary the overlap integrals for both transitions to have the same (opposite) sign. For the states $\Phi^{\Pi_u}_{60}, \Phi^{\Pi_u}_{61}$ the overlaps for the upwards (downwards) transition are both positive (negative) [see also SM~\cite{supp}]. This leads to enhanced FC factor for $\Psi_-$ in both transitions.

\textit{Photoassociation in the Cavity.}---The phenomenon of FC transfer around the avoided crossing and the selective enhancement of polariton states suggests that molecular photoassociation of $\textrm{Rb}_2$ molecules to the $\Sigma^+_g$ PES  can be controlled with resonant light-molecule coupling. To demonstrate this we perform a prototype STIRAP photoassociation for $\textrm{Rb}_2$ in the cavity, employing the $\Lambda$-transition scheme with Gaussian laser fields shown in Fig.~\ref{Setup}(c). The first laser drives the transition $\Sigma^+_{u} \leftrightarrow \Pi^+_u$, and the second one the $\Pi_u \leftrightarrow \Sigma^+_g$. The populations of each PES during the STIRAP obey the following coupled differential equations~\cite{DenschlagPRL},
\begin{eqnarray}\label{stirap}
    \textrm{i} \dot{\Sigma}_u(t)&=&-\textrm{i}\frac{\gamma_{\Sigma_u}}{2} \Sigma_u(t) -\frac{\Omega_1}{2}\Pi_u(t),\nonumber\\
    \textrm{i} \dot{\Pi}_u(t)&=&-\textrm{i}\frac{\gamma_{\Pi_u}}{2}\Pi_u(t) -\frac{1}{2}\left(\Omega_1\Sigma_u(t)+\Omega_2\Sigma_g(t)\right),\\
    \textrm{i}\dot{\Sigma}_g(t)&=&-\textrm{i}\frac{\gamma_{\Sigma_g}}{2}\Sigma_g(t) -\frac{\Omega_2}{2}\Pi_u(t).\nonumber
\end{eqnarray}
Here, $\Omega_{1,2}$ are the Rabi frequencies corresponding to the two laser fields $\mathbf{E}_{1,2}$ respectively. Further, we use the decay rates $\gamma_{\Sigma_u}=2\pi\times0.72\;\rm{kHz}$, $\gamma_{\Pi_u}=2\pi\times12\;\rm{MHz}$ for the $\Pi_u$ and the $\Sigma_u$ states respectively taken from the experiment of Ref.~\cite{DenschlagPRL}. Notice that $\gamma_{\Sigma_g}=0$, because we aim at the vibrational ground state and thus there is no lower state for the molecule to decay to. Also, in accordance with the theoretical modeling of the experiment of Ref.~\cite{DenschlagPRL}, the detunings of the lasers are assumed to be zero. Within the Born-Oppenheimer approximation the Rabi frequencies $\Omega_{1,2}$ for the laser fields $\mathbf{E}_{1,2}$ are a product of the FC factors of the vibrational levels and the electronic dipole moment for the transition $5s \rightarrow 5p$. Therefore, it holds that $\Omega_{i} = \frac{|\mathbf{E}_{i}|}{\hbar}\langle \Phi_{m}|\Phi_{n} \rangle  \langle 5s|d_e|5p\rangle$. The value for the electronic transition $\langle 5s|d_e|5p\rangle $ is obtained from the $A_{5s5p}=3.812\times 10^7 \textrm{s}^{-1}$ Einstein coefficient~\cite{scully1997} provided in Ref.~\cite{NIST_ASD}. The decisive difference for the STIRAP in the cavity is that the bare FC factors of $\textrm{Rb}_2$ are replaced by the FC factors of the polariton states. 

\begin{figure}[H]
     \centering
     \begin{subfigure}[b]{0.5\textwidth}
         \centering
        \includegraphics[width=\columnwidth]{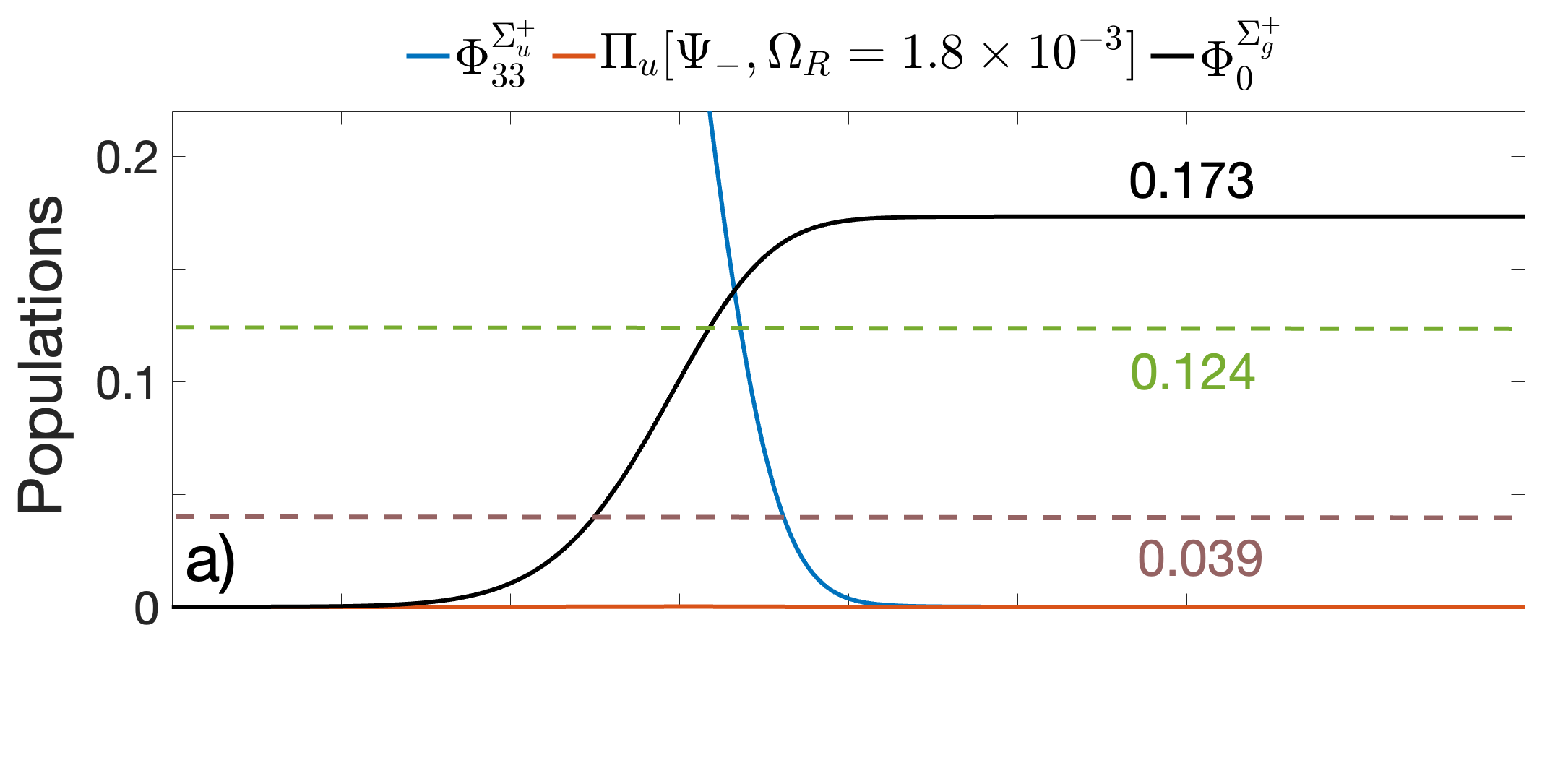}
        \vspace{-1.1cm}
   \end{subfigure}
   \begin{subfigure}[b]{0.5\textwidth}
         \centering
         \includegraphics[width=\columnwidth]{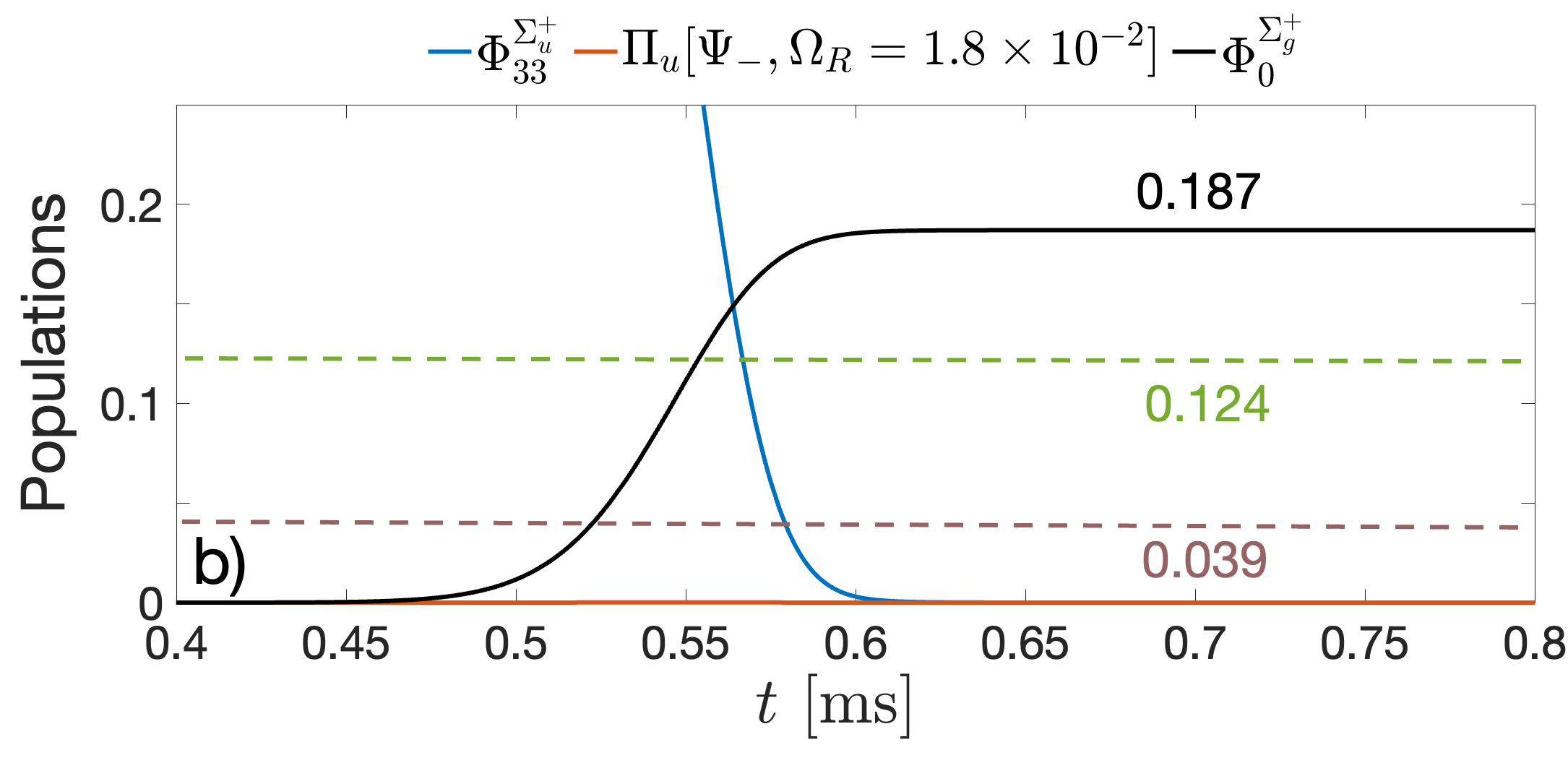}
         \end{subfigure}
    \caption{STIRAP photoassociation to $\Phi^{\Sigma^+_g}_0$ through the lower polariton state $\Psi_-$ at the exact resonance point. In a) for weak coupling $\Omega_R=1.8\times10^{-3}$ the photoassociation is $17.3\%$ and in b) for strong coupling $\Omega_R=1.8\times10^{-2}$ reaches $18.7\%$. As compared to photoassociation through the molecular states $\Phi^{\Pi_u}_{60}$ (green dashed line) and the states $\Phi^{\Pi_u}_{61}$ (brown dashed line) the lower polariton demonstrates enhanced molecular photoassociation. This is a consequence of the enhanced polariton FC factors. }
    \label{STIRAP}
       \end{figure}

The time-evolution of the populations of each state following STIRAP are illustrated in Figs.~\ref{STIRAP}(a) and (b) for the lower polariton branch for weak ($\Omega_R=1.8\times 10^{-3}$) and strong ($\Omega_R=1.8\times10^{-2}$) coupling respectively. In both cases, as expected after the application of the second pulse, around $t=1ms$, the system attains a steady state where the population of the molecular ground state $\Phi^{\Sigma^+_g}_0$ becomes stationary because the system is no longer driven. Most importantly, we find that the population of molecules in the vibrational ground state, $\Phi^{\Sigma^+_g}_0$, is substantially enhanced when STIRAP is performed through the lower polariton $\Psi_-$. Particularly, for weak coupling $\Omega_R=1.8\times10^{-3}$ photoassociation of molecules becomes $17.3\%$ and for strong coupling $\Omega_R=1.8\times10^{-2}$ reaches $18.7\%$. In comparison to molecular photoassociation through the bare vibrational states $\Phi^{\Pi_u}_{60}$ and $\Phi^{\Pi_u}_{61}$ which are $12.4\%$ and $3.9\%$ respectively (see the dashed lines in Fig.~\ref{STIRAP}) we deduce that with respect to the state $\Phi^{\Pi_u}_{60}$ the relative percentage increase is $39.5\%$ ($50.8\%$) for $\Omega_R=1.8\times10^{-3}$ ($\Omega_R=1.8\times10^{-2}$). This is a consequence of the enhanced FC factors of the lower polariton state shown in Fig.~\ref{PFCs}. For completeness we mention that the molecular photoassociation through the upper polariton does not exhibit any enhancement but it is extremely suppressed, see more details in SM~\cite{supp}. This happens because the FC factors of $\Psi_+$ decrease drastically at the resonance [see Figs.~\ref{PFCs}(a) and (b)].

\textit{Discussion and Outlook.}---The photoassociation of ultracold molecules can be substantially enhanced by resonant vibrational strong coupling to the vacuum cavity. This is due to the phenomenon of polariton FC transfer that emerges around the avoided crossings between the polaritons. This generic mechanism, arising from the photon-molecule entanglement around a resonance, can be understood within the standard JC paradigm. In the strong coupling regime, however, the JC model is not adequate and the full length gauge Hamiltonian is necessary for the accurate description of the photon-molecule interaction. The relative percentage increase of molecular photoassociation is substantial, and in the strong coupling regime, it approaches  $\sim 50\%$.  

It is important to mention however, that the photoassociation through polaritonic states is not always advantageous for all vibrational states. The polaritonic STIRAP demonstrates clear advantage only when the FC factors for both branches in the same polariton state are enhanced in unison in both directions of the $\Lambda$-scheme. Thus, observing an avoided crossing does not directly imply that a chemical process of interest will be necessarily modified; for a more detailed discussion, see the SM~\cite{supp}.

This work introduces the element of cavity polaritonic control to ultracold chemistry, and paves the way for future investigations at the intersection between cQED and ultracold chemistry. A particularly exciting research direction is to couple dipolar ultracold molecules, e.g. $\rm{KRb}$, to cavity fields where due to the existence of a permanent dipole moment, the ground state of the molecule can be modified through polariton formation~\cite{RokajHarmonic}. In addition, $\rm{KRb-KRb}$ collisions~\cite{NiColdCollisions} inside a cavity can be investigated. In the many-body regime of a gas of ultracold molecules~\cite{zhang2021transition,li2021tuning, Brantut2021}, coherent control of molecular orientation by the cavity field is an enticing possibility~\cite{sidler2023unraveling}. Along these lines, recently, universal  behavior has been reported for pair-polariton states emerging when a Fermi gas is coupled to a cavity~\cite{Brantut2021}. The unparalleled precision and controlability provided in ultracold physics, can aid to uncover mechanisms behind the effects observed in polaritonic chemistry. Achieving mechanistic understanding of these phenomena through ultracold molecules will be an important leap forward.

\begin{acknowledgements}
The authors thank Dan Stamper-Kurn and Giacomo Valtolina for stimulating discussions. We acknowledge support from the NSF through a grant for ITAMP at Harvard University.
\end{acknowledgements}

\bibliography{Bibliography}

\clearpage
\pagebreak

\onecolumngrid
\widetext
\setcounter{figure}{0}
\setcounter{equation}{0}
\renewcommand{\theequation}{S\arabic{equation}}
\renewcommand{\thefigure}{S\arabic{figure}}
\begin{center}
\textbf{ SUPPLEMENTARY MATERIAL} \\

\end{center}

\section{Molecular Model}

Here, we provide details about the potential energy surfaces (PESs) used for $\textrm{Rb}_2$, the respective vibrational wavefunctions involved in the STIRAP photoassociation, as well as the bare Franck-Condon (FC) factors. For the description of the $\rm{Rb}_2$ we employ the well-known Morse model~\cite{Morse}
\begin{eqnarray}
    \mathcal{H}_{M}=-\frac{\hbar^2 }{2\mu}\frac{\partial^2}{\partial R^2}+D\left[e^{-2w(R-R_0)}-2e^{-w(R-R_0)}\right],  
\end{eqnarray}
 where $\mu$ is the reduced mass of the dimer, $D$ is the depth of the potential well, $w$ the width of the potential, and $R_0$ denotes the equilibrium bond distance~\cite{Morse}. To perform the photoassociation process in the main text we used the following PESs of $\rm{Rb}_2$: $\Sigma^+_u$, $\Pi_u$ and $\Sigma^+_g$. The PESs are modeled with the Morse potentials and its corresponding parameters were obtained from the experimental data of Ref.~\cite{RB2potentials}. The explicit characteristics of these PESs are given in Table~\ref{Table1}.

\begin{center}\label{Table1}
    \begin{table}[h!]
\begin{tabular}{c  c  c  c c c} 
 \hline
 \hline
 \\ 
 State \; & \; $D[\textrm{cm}^{-1}]$ \; & \; $\omega_e[\textrm{cm}^{-1}]$ \; & \; $R_0\left[\textrm{\AA}\right]$ &\; $T[\textrm{cm}^{-1}]$\; & Dissociation limits \\ \\
 \hline
 \hline\\
 $(1)^3\Sigma^+_u$ & 240 & 14 & 6.13 & 3587 & 5s+5s \\ 
 \\\hline\\
$(1)^3\Pi_u$ & 6828 & 59 & 4.20 & 9779 & 5s+5p\\ [1ex] 
\\\hline\\
$X^1\Sigma^+_g$ & 3827 & 59 & 4.23 & 0 & 5s+5s\\ [1ex]
 \hline
 \hline
\end{tabular}
\caption{Experimentally obtained parameters for the Morse potentials describing the potential energy surfaces of the $\textrm{Rb}_2$ molecule as reported in Ref~\cite{RB2potentials}. Note that $\omega_e=w\sqrt{2D/\mu}$.}
\end{table}
\end{center}

\begin{figure}[H]
\centering
\begin{subfigure}[b]{0.3\textwidth}
\centering
\includegraphics[width=\columnwidth]{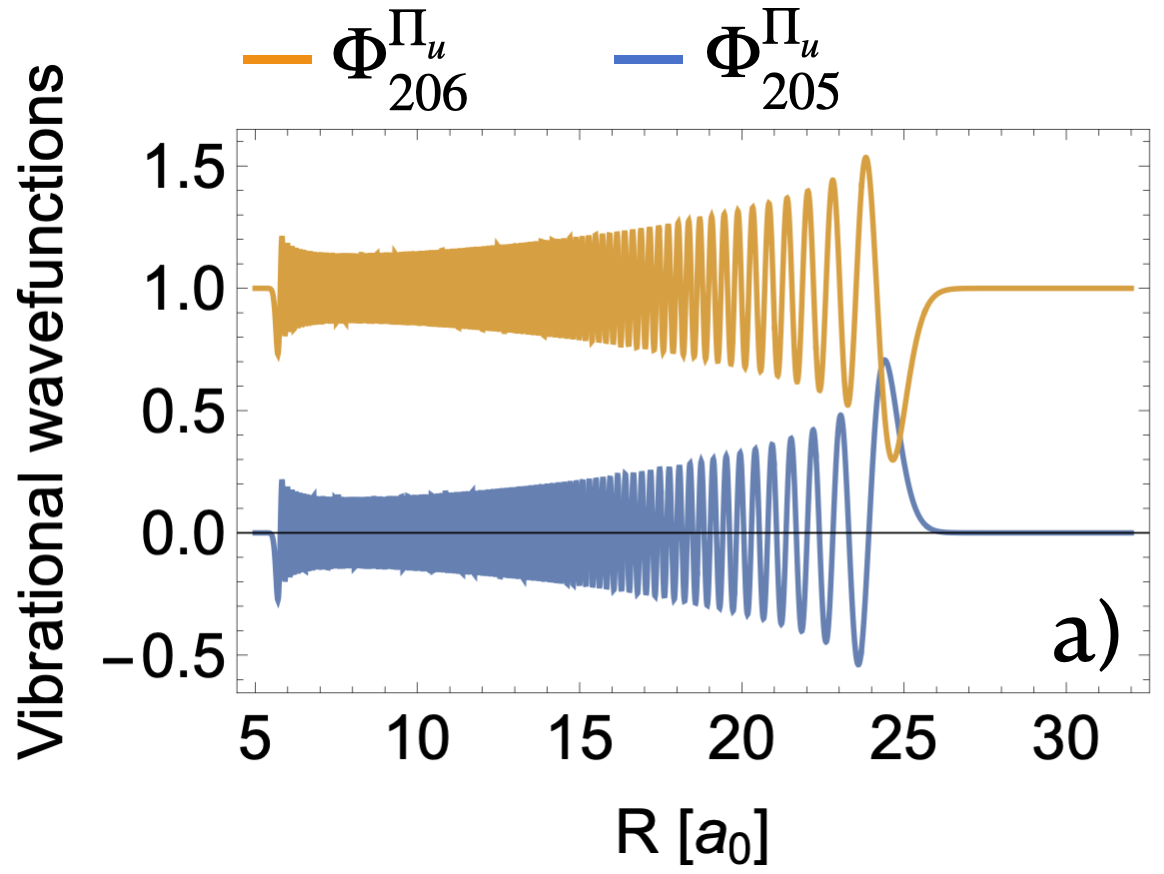}
\end{subfigure}
\hfill
\begin{subfigure}[b]{0.3\textwidth}
\centering
\includegraphics[width=0.95\columnwidth]{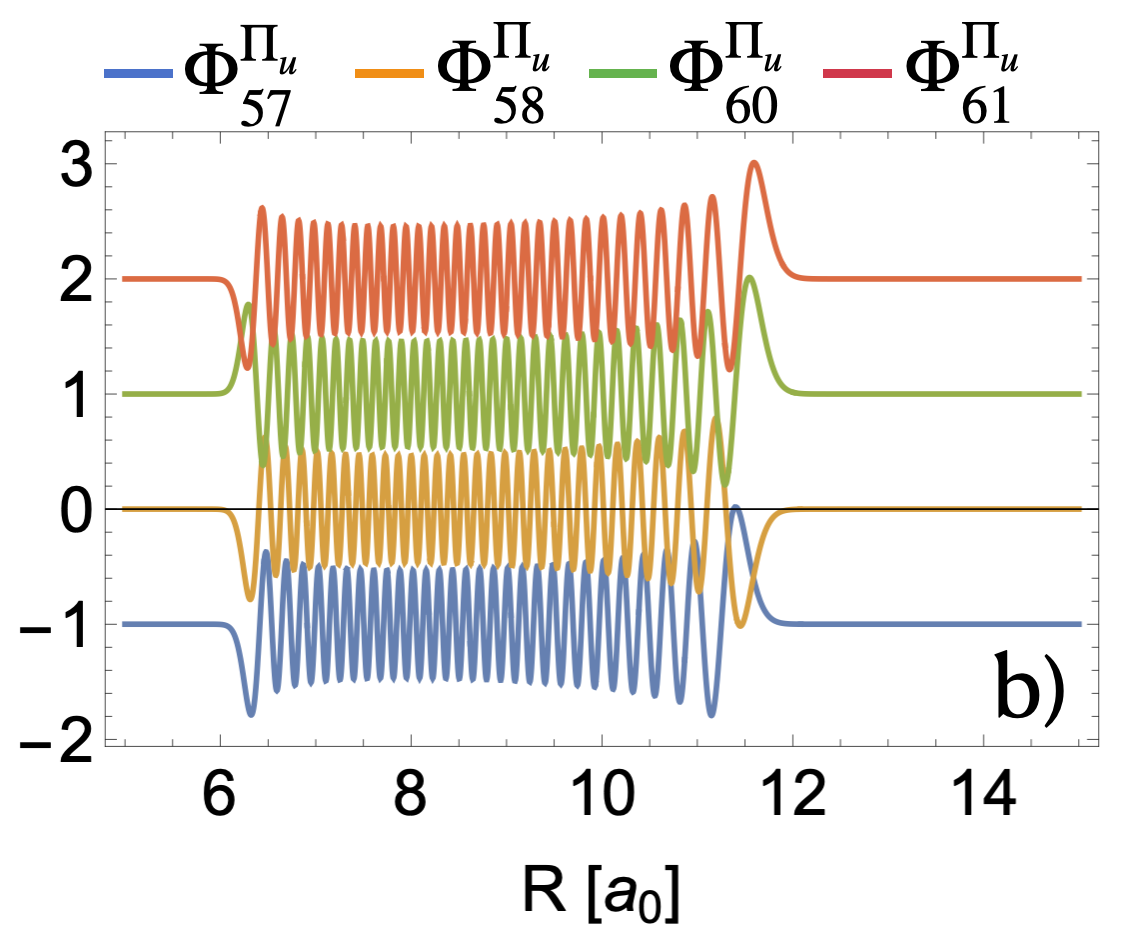}
\end{subfigure}
\hfill
\begin{subfigure}[b]{0.3\textwidth}
\centering
\includegraphics[width=0.95\columnwidth]{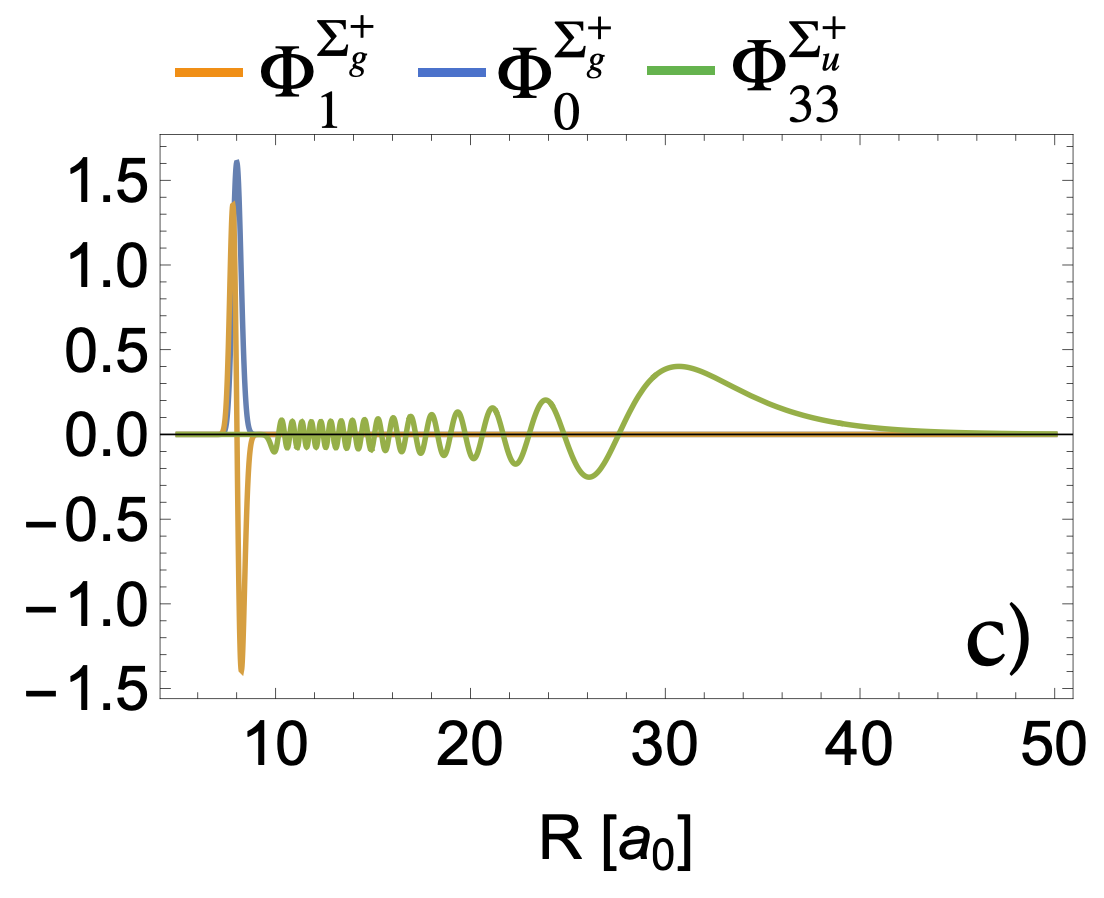}
\end{subfigure}
\caption{Vibrational wavefunctions of all the states considered in the main text and in supplement (see legends). Several of the vibrational wavefunctions are shifted such that all profiles are well discernible.}
\label{VibStates}
\end{figure}

The vibrational states were obtained through numerical diagonalization. Characteristic examples of vibrational wavefunctions to explore photoassociation are shown in Fig.~\ref{VibStates}. Particularly, in the main text, the states $\Phi^{\Pi_u}_{60},\Phi^{\Pi_u}_{61}, \Phi^{\Sigma^+_u}_{33} $ and $\Phi^{\Sigma^+_g}_{0}$ presented in Figs.~\ref{VibStates}(b), (c) have been employed.

\begin{figure}[H]
\centering
\begin{subfigure}[b]{0.49\textwidth}
\centering
\includegraphics[width=\columnwidth]{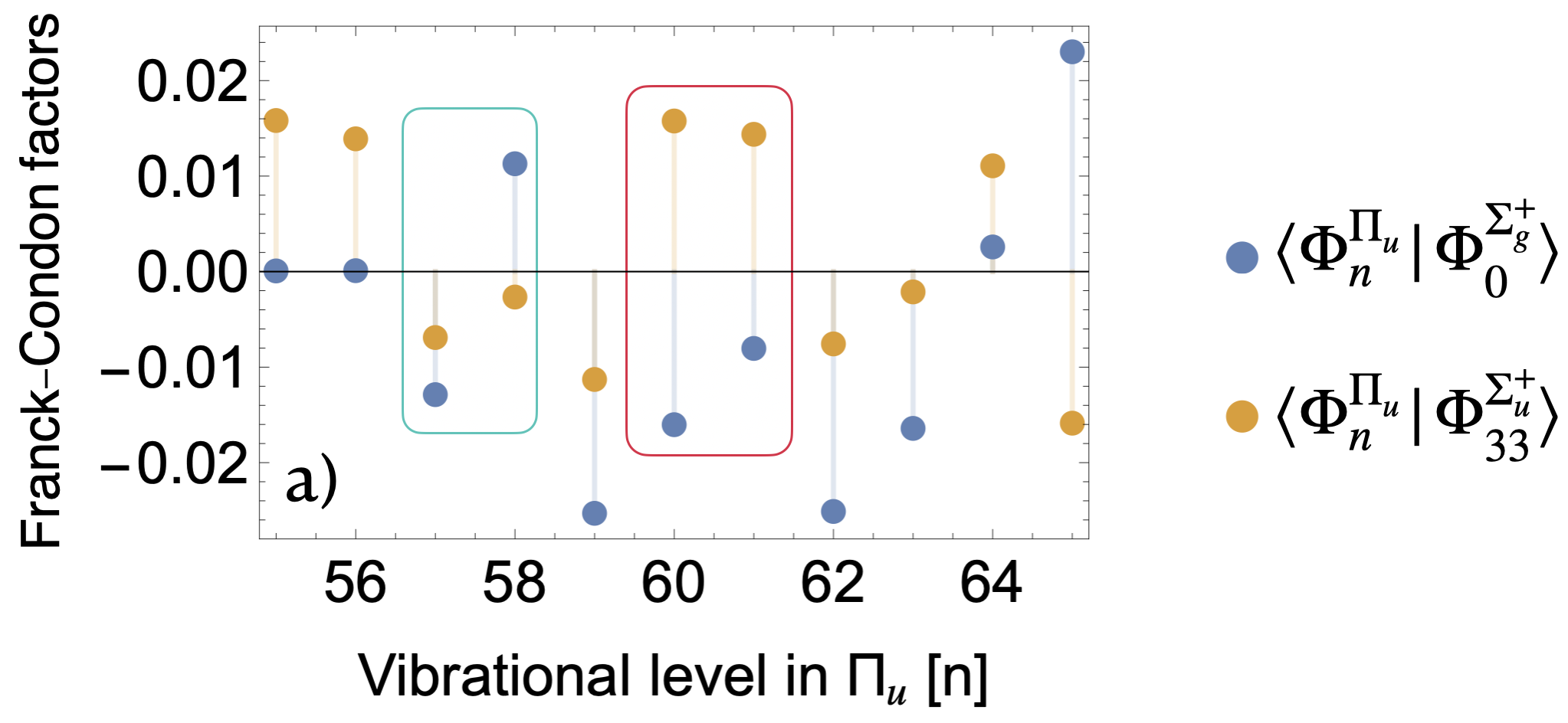}
\end{subfigure}
\hfill
\begin{subfigure}[b]{0.49\textwidth}
\centering
\includegraphics[width=0.95\columnwidth]{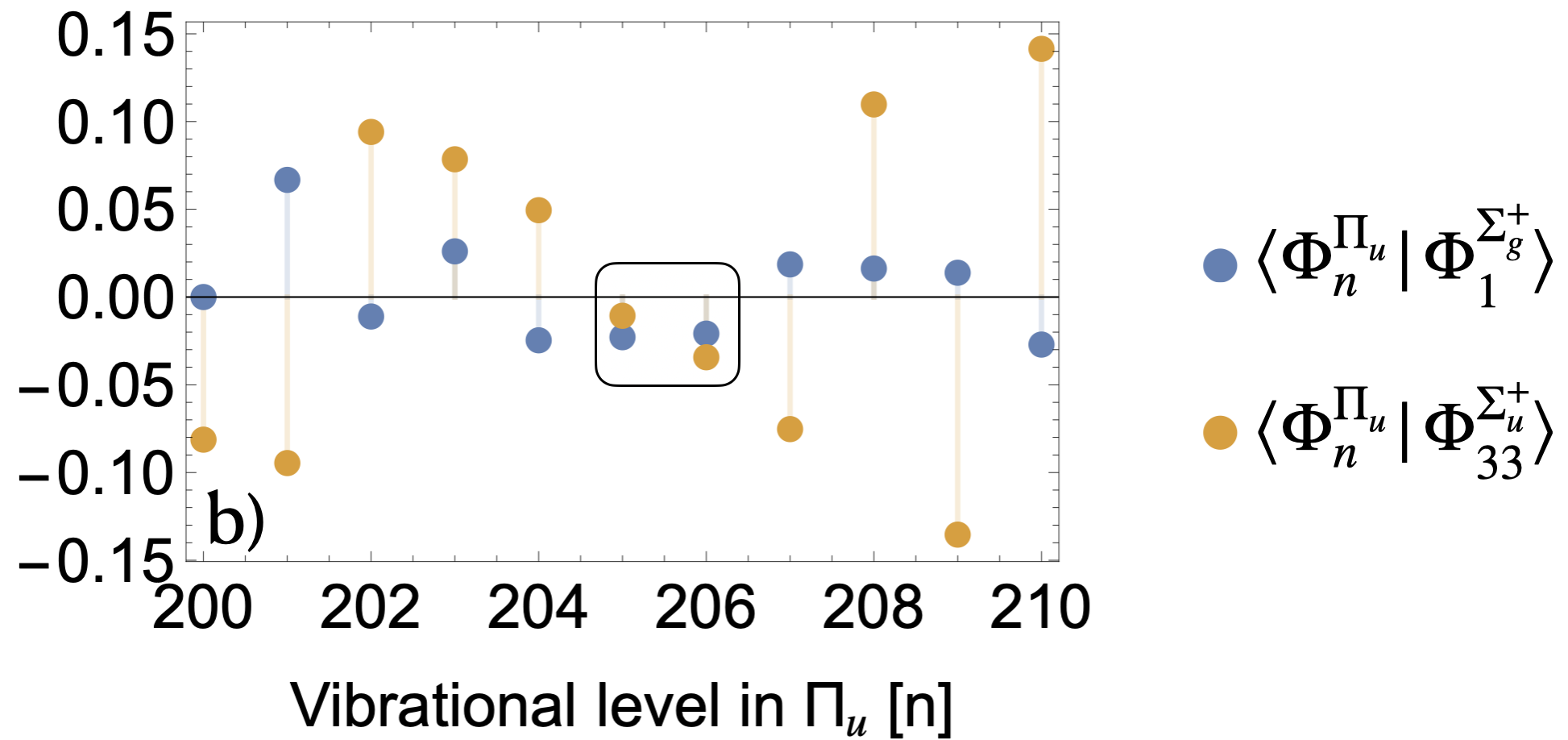}
 \end{subfigure}
\caption{FC factors of different vibrational states in the $\Pi_u$ PES (see legend). The transitions to a) $\Phi^{\Sigma^+_g}_0$ and b) $\Phi^{\Sigma^+_g}_1$ are shown. In both panels the upwards transitions from $\Phi^{\Sigma^+_u}_{33}$ are provided which are used in the $\Lambda$-scheme for photoassociation. }
\label{OVerlaps}
\end{figure}

Further, in Fig.~\ref{OVerlaps} we show several of the bare FC factors for the in $\Pi_u$ PES, among which in panel (a) are the states $\Phi^{\Pi_u}_{60}, \Phi^{\Pi_u}_{61} $ which we used for the STIRAP photoassociation in the main text. As it can be seen from Fig.~\ref{OVerlaps}(a) the FC factors of $\Phi^{\Pi_u}_{60}$ and $\Phi^{\Pi_u}_{61}$ (in the red box) for both the upwards (from $\Phi^{\Sigma^+_u}_{33}$) and the downwards (to $\Phi^{\Sigma^+_g}_{0}$) transitions have the same sign. This explains the fact that the corresponding FC factors of the lower polariton for both transitions are enhanced. This leads subsequently to the enhancement of photoassociation of molecules through the lower polariton branch as it was discussed in the main text. Moreover, in Fig.~\ref{OVerlaps}(a) we present the FC factors of the states $\Phi^{\Pi_u}_{57}$ and $\Phi^{\Pi_u}_{58}$ (in the blue box) which are used for the photoassociation in Sec.~\ref{No advantage}. Finally, in Fig.~\ref{OVerlaps}(b) we provide additionally some of the FC factors of higher lying vibrational states for the upwards (from $\Phi^{\Sigma^+_u}_{33}$) and downwards (to $\Phi^{\Sigma^+_g}_{1}$) transitions. In the black box we highlight the FC factors of the states $\Phi^{\Pi_u}_{205}$ and $\Phi^{\Pi_u}_{206}$ which are used in Sec.~\ref{second stirap}.

\section{Molecule-Photon Hamiltonian}
The coupled molecule-photon system is described by the Pauli-Fierz Hamiltonian in the length gauge~\cite{rokaj2017, cohen1997photons}
\begin{eqnarray}
    \mathcal{H}=\mathcal{H}_M-\frac{\hbar \omega}{2}\frac{\partial^2}{\partial q^2} +\frac{\hbar \omega}{2}\left(q-\frac{\lambda d_R}{\sqrt{\hbar \omega}}\right)^2.
\end{eqnarray}
The cavity field is described by the displacement coordinate $q$ and its conjugate momentum $\partial/\partial q$ while the cavity frequency is $\omega=\pi c/L$. Equivalently, the cavity mode can be described in terms of annihilation and creation operators $a=(q+\partial_q)/\sqrt{2},\; \textrm{and} \; a^{\dagger}=(q-\partial_q)/\sqrt{2}$, which satisfy bosonic commutations relations $[a,a^{\dagger}]=1$. After expanding the quadratic term $\left(q-\frac{\lambda d_R}{\sqrt{\hbar \omega}}\right)^2$ and formulating the Hamiltonian in terms of annihilation and creation operators we have
\begin{eqnarray}
\mathcal{H}=\mathcal{H}_M+\hbar\omega\left(a^{\dagger}a+\frac{1}{2}\right)-\sqrt{\frac{\hbar \omega}{2}}\lambda d_R\left(a+a^{\dagger}\right)  +\frac{\lambda^2 d^2_R}{2}.
\end{eqnarray}
The cavity field strength $\lambda=1/\sqrt{\epsilon_0\mathcal{V}}$ depends on the effective cavity volume $\mathcal{V}=AL$ which is a product of the mirror distance $L$ and the area of the cavity mirrors $A$. The cavity frequency $\omega=\pi c/L$ and the mirror distance $L$ are also related, and consequently the field strength $\lambda$ depends on $\omega$ as well, i.e. $\lambda=\sqrt{\omega/\epsilon_0\pi c A}$. Further, $c$ is the speed of light and $\epsilon_0$ the vacuum permittivity. We use the linear dipole moment $d_R=\mu_0 R$ with strength $\mu_0$, and  express the molecule-photon $\mathcal{H}$ in terms of the light-matter coupling constant $g$ which is independent of the cavity frequency,
\begin{eqnarray}
\mathcal{H}=\mathcal{H}_M+\hbar\omega\left(a^{\dagger}a+\frac{1}{2}\right)-\omega g\sqrt{\frac{\hbar}{2}}R\left(a+a^{\dagger}\right)  +\frac{\omega g^2}{2}R^2, \;\; \textrm{where}\;\; g=\frac{\mu_0\lambda}{\sqrt{\omega}}=\frac{\mu_0}{\sqrt{A\pi c\epsilon_0}}. 
\end{eqnarray}
To treat the interacting molecule-photon coupled system we expand the Hamiltonian in the basis consisting of the molecular eigenstates of the $\Pi_u$ PES $\{|\Phi^{\Pi_u}_n\rangle\}$ and the eigenstates $\{|i\rangle\}$ of the bosonic cavity mode~\cite{GriffithsQM}. In this way, we obtain the matrix form $\mathcal{H}^{ji}_{mn}:=\langle j|\langle \Phi^{\Pi_u}_m|\mathcal{H}|\Phi^{\Pi_u}_n\rangle|i\rangle$ of the Hamiltonian $\mathcal{H}$ which reads
\begin{eqnarray}
    \mathcal{H}^{ji}_{mn}=\left[\mathcal{E}_n+ \hbar\omega\left(i+\frac{1}{2}\right)\right]\delta_{mn}\otimes\delta_{ji} - \omega g\sqrt{\frac{\hbar}{2}}\langle \Phi^{\Pi_u}_m|R|\Phi^{\Pi_u}_n\rangle\otimes\left[\sqrt{i}\delta_{j,i-1}+\sqrt{i+1}\delta_{j,i+1}\right]+\frac{\omega g^2}{2}\langle\Phi^{\Pi_u}_m| R^2|\Phi^{\Pi_u}_n\rangle\otimes \delta_{ji}.\nonumber\\
\end{eqnarray} 
We note that $\mathcal{E}_n$ are the eigenenergies of the molecular Hamiltonian $\mathcal{H}_M$ described by the Morse model.

\section{Additional Details of STIRAP to $\Phi^{\Sigma^+_g}_{0}$  }

\begin{figure}[H]
\centering
\begin{subfigure}[b]{0.49\textwidth}
\centering
\includegraphics[width=\columnwidth]{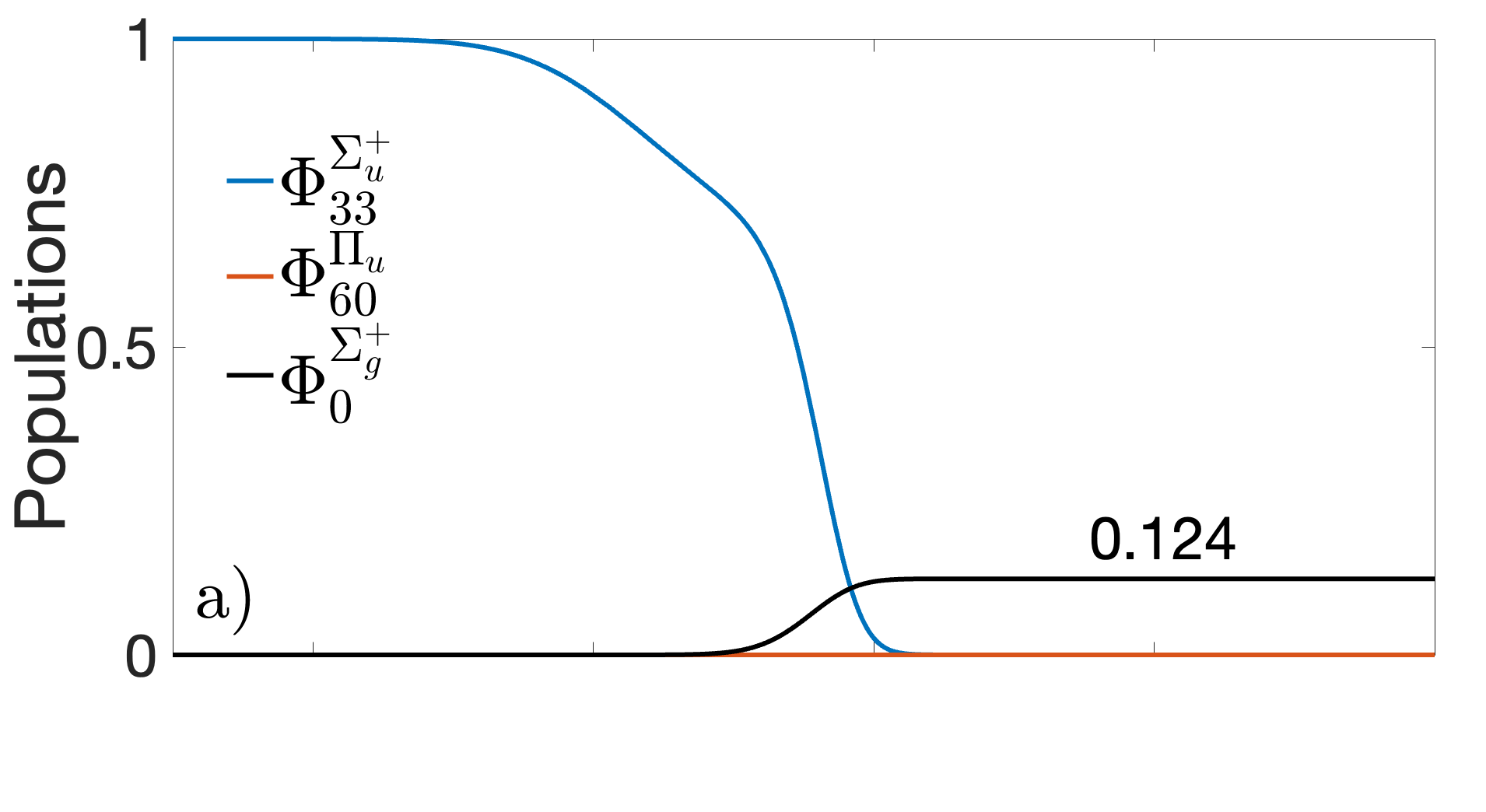}
\end{subfigure}
\hfill
\begin{subfigure}[b]{0.49\textwidth}
\centering
\includegraphics[width=\columnwidth]{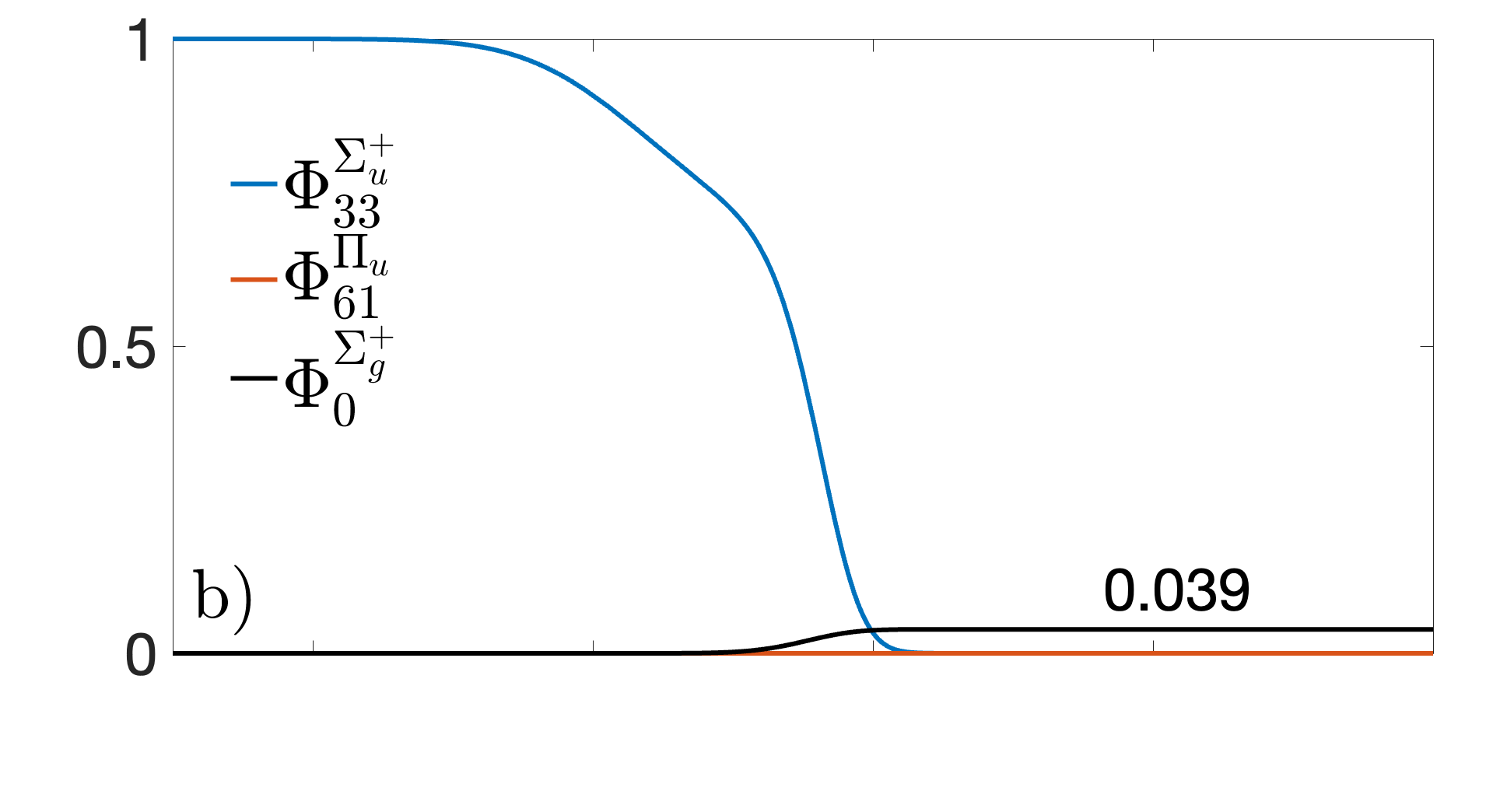}
\end{subfigure}
\hfill
\begin{subfigure}[b]{0.49\textwidth}
\centering
\includegraphics[width=\columnwidth]{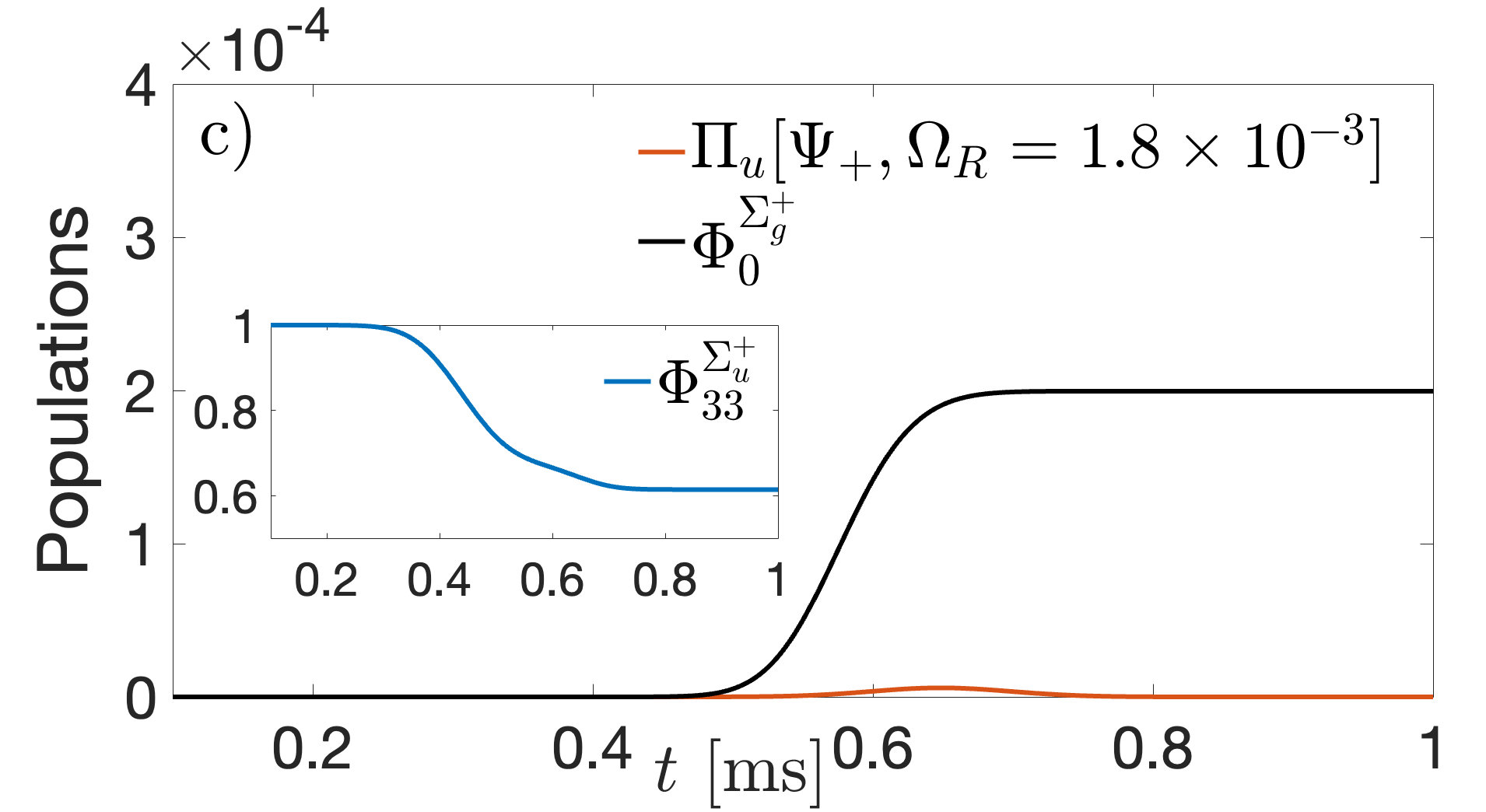}
\end{subfigure}
\hfill
\begin{subfigure}[b]{0.49\textwidth}
\centering
\includegraphics[width=\columnwidth]{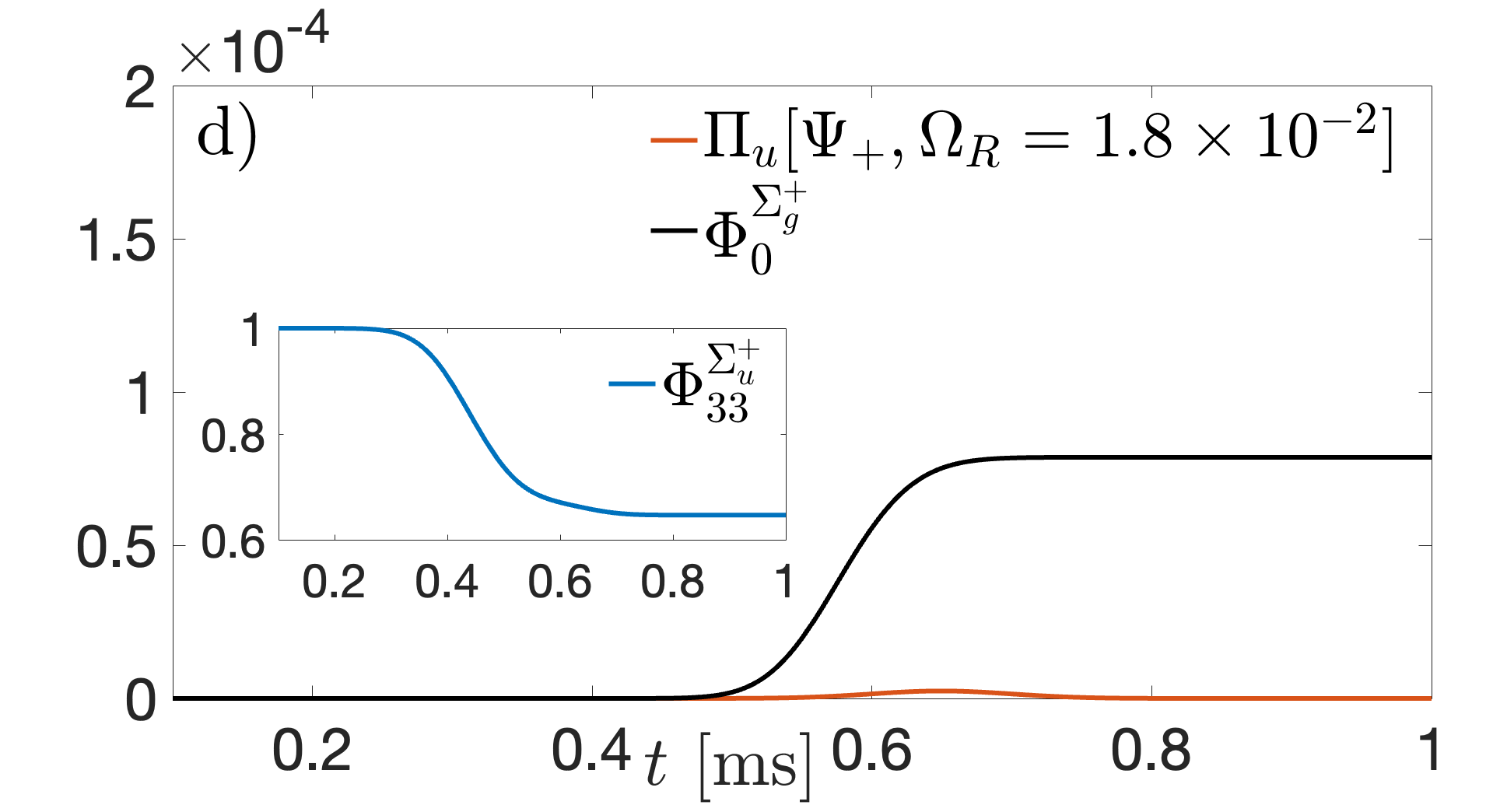}
\end{subfigure}
\caption{Full time-evolution of the populations of the three states (see legend) involved in the STIRAP. The STIRAP for a) $\Phi^{\Pi_u}_{60}$ and b) $\Phi^{\Pi_u}_{61}$ is shown. The STIRAP for the upper polariton $\Psi_+$ for c) $g=0.001$ and d) $g=0.01$ is demonstrated. The time evolution of the population during STIRAP to $\Phi^{\Sigma^+_g}_{0}$ through the upper polariton is also depicted. The upper polariton features very low photoassociation of $\textrm{Rb}_2$ molecules due to the very strongly suppressed FC factors in both transitions. }
\label{UpperStirap}
\end{figure}

Let us now provide some additional data and supporting computations, about the photoassociation of $\textrm{Rb}_2$ molecules to the vibrational ground state $\Phi^{\Sigma^+_g}_{0}$ presented in the main text. Recall that in the main text we only discussed the STIRAP to $\Phi^{\Sigma^+_g}_0$ through the lower polariton state $\Psi_-$ which exhibits the enhanced FC factors around the avoided crossing. For completeness, in Fig.~\ref{UpperStirap} we provide the molecular yield when STIRAP is performed through bare molecular states $\Phi^{\Pi_u}_{60}$ [Figs.~\ref{UpperStirap}(a)] and $\Phi^{\Pi_u}_{61}$ [Figs.~\ref{UpperStirap}(a)]. Further, the STIRAP through the upper polariton state $\Psi_+$, exhibiting  suppressed FC factors, for $\Omega_R=1.8\times 10^{-3}$ and $\Omega_R=1.8\times 10^{-2}$ is illustrated in  Figs.~\ref{UpperStirap}(c) and (d). In all cases, initially ($t=0$) the state $\Phi^{\Sigma^+_u}_{33}$ is fully populated. However, as time-evolves the population of the state $\Phi^{\Sigma^+_u}_{33}$ decreases mostly due to the decay term $\gamma_{\Sigma_u}$ of the $\Sigma_u$ state, see also Eq.~\ref{stirap} in the main text. Specifically, when the Gaussian pulses are applied [Fig.~\ref{Setup}(c) in the main text] population is transferred from the initial state $\Phi^{\Sigma^+_u}_{33}$ to the final one $\Phi^{\Sigma^+_g}_{0}$. For STIRAP through $\Phi^{\Pi_u}_{60}$ [Fig.~\ref{UpperStirap} (a)] the amount of $12.4\%$ of rubidium atoms turn into molecules, while for the $\Phi^{\Pi_u}_{61}$ [Fig.~\ref{UpperStirap}(b)] we observe only $3.9\%$ production of $\textrm{Rb}_2$. For STIRAP through the upper polariton $\Psi_+$ the molecular yield is extremely suppressed to $\sim 0.01\%$ both for weak [Fig.~\ref{UpperStirap}(c)] and strong [Fig.~\ref{UpperStirap}(d)] coupling, as expected from the substantially suppressed FC factors given in the main text [Fig.~\ref{PFCs}].

\subsection{Comparison to the Jaynes-Cummings model}

\begin{figure}[H]
     \centering
     \begin{subfigure}[b]{0.45\textwidth}
         \centering
        \includegraphics[width=\columnwidth]{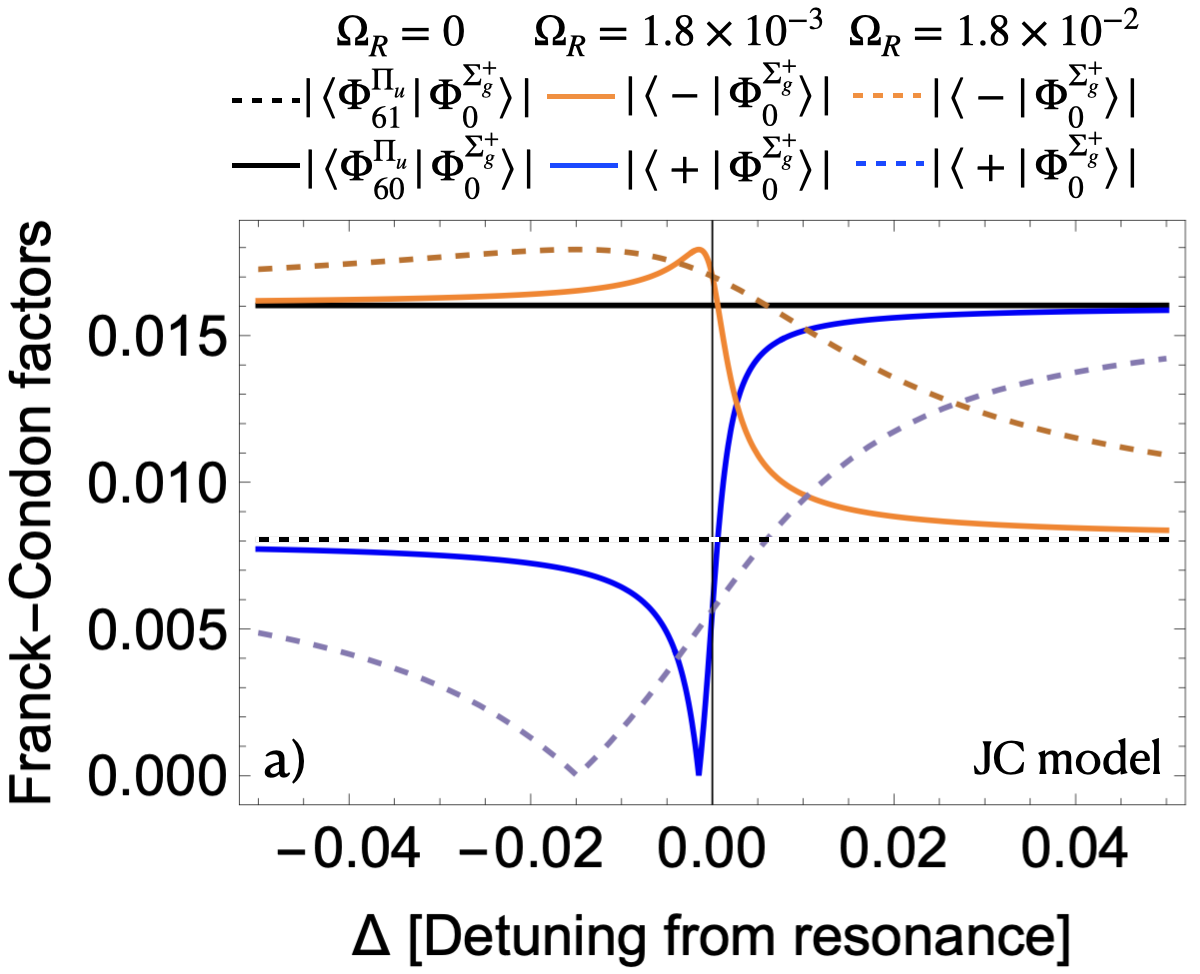}
   \end{subfigure}
     \hfill
     \begin{subfigure}[b]{0.45\textwidth}
         \centering
         \includegraphics[width=0.95\columnwidth]{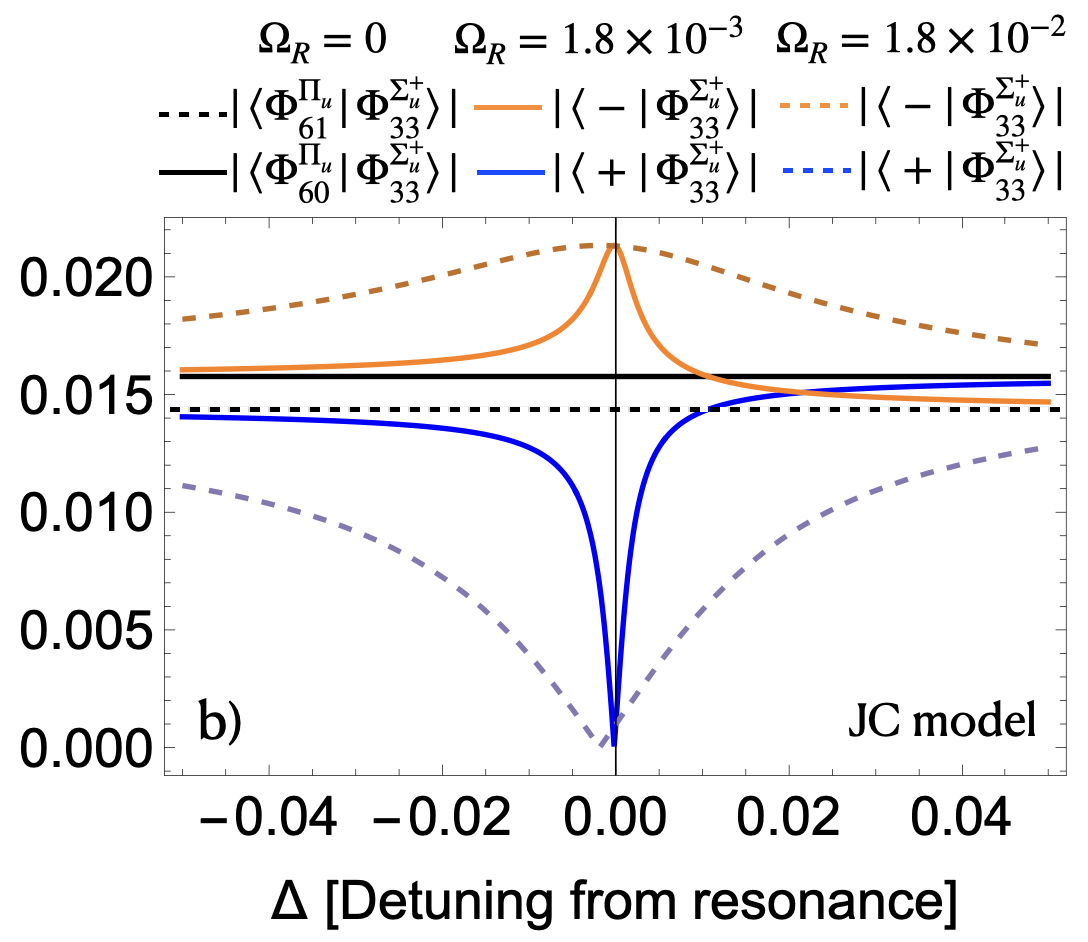}
 \end{subfigure}
\caption{Bare and polaritonic FC factors of $\rm{Rb}_2$ for $\Omega_R=1.8\times10^{-3}$ and $\Omega_R=1.8\times 10^{-2}$ computed within the JC model. a) FC factors for the downwards transition to $\Phi^{\Sigma^+_u}_{0}$ from $\Phi^{\Pi_u}_{60},\Phi^{\Pi_u}_{61}$ and the respective JC states $|\pm\rangle$ (see legend). b) FC factors regarding the upward transition from $\Phi^{\Sigma^+_u}_{33}$ to $\Phi^{\Pi_u}_{60},\Phi^{\Pi_u}_{61}$ and the JC states $|\pm\rangle$ (see legend). In both panels at resonance $|-\rangle$ shows enhanced FC factors, while the FC factors of $|+\rangle$ are suppressed. The JC model for weak coupling $\Omega_R=1.8\times 10^{-3}$ is in agreement with the full Hamiltonian [Eq.~(\ref{Hamiltonian})]. However, the JC model for strong coupling $\Omega_R=1.8\times 10^{-2}$ disagrees with the full model as no further effect on the FC factors is observed at the resonance point.}
\label{JCFCfactors}
\end{figure}

In the main text we discussed that in the weak coupling regime, $\Omega_R=1.8\times 10^{-3}$, the hybridization between the vibrational states and the cavity photons, but also the corresponding FC factors of the polariton states can be captured using the Jaynes-Cummings model~\cite{Vogel, scully1997}. Indeed, for the case where the interaction term between the two-level system and light is $\mathcal{H}_{\rm{int}}=-(\hbar \Omega_R/2)\sigma^+\hat{a}+\textrm{h.c.}$ it can be proven that the JC states are~\cite{Vogel}
\begin{eqnarray}\label{JCstates}
    |+\rangle=\alpha_+|\Phi^{\Pi_u}_n\rangle|1\rangle-\alpha_-|\Phi^{\Pi_u}_{n+1}\rangle|0\rangle\;\textrm{and}\;
    |-\rangle=\alpha_-|\Phi^{\Pi_u}_n\rangle|1\rangle+\alpha_+|\Phi^{\Pi_u}_{n+1}\rangle|0\rangle\; \textrm{where}\; \alpha_{\pm}=\frac{1}{\sqrt{2}}\sqrt{1\pm \frac{\Delta}{\sqrt{\Delta^2+\Omega^2_R}}}.
\end{eqnarray}
Here, $\Delta$ is the detuning between the cavity mode and the two-level transition. The minus sign in the interaction term of the JC model stems from the form of the length gauge Hamiltonian and the fact that the transition dipole matrix element is positive in our example $\langle \Phi^{\Pi_u}_{60}|R|\Phi^{\Pi_u}_{61}\rangle >0$. Using the analytic expression for the JC states in Fig.~\ref{JCFCfactors} we show the FC factors of the JC states $|\pm\rangle$ for the two values of the light-matter coupling strength and as a function of the detuning $\Delta$. The FC factors for the upwards transition from $\Phi^{\Sigma^+_u}_{33}$ to the states in the $\Pi_u$ PES are illustrated in Fig.~\ref{JCFCfactors}(a), while the FC factors for the downwards transition to the vibrational ground state $\Phi^{\Sigma^+_g}_0$ are provided in Fig.~\ref{JCFCfactors}(b). For both transitions we observe that the FC factor of the lower polariton is enhanced around the resonance point $\Delta=0$, while the FC factor of the upper polariton is suppressed. This behavior is in agreement with the polaritonic FC factors obtained from the full length gauge Hamiltonian [Eq.~(\ref{Hamiltonian}) in the main text] which includes the counter-rotating terms and the dipole self-energy. For the stronger light-matter coupling where $\Omega_R=1.8\times 10^{-2}$,  however, the JC model fails to capture the further modification of the FC factors at the resonance point ($\Delta=0$). The JC states at $\Delta=0$ are independent of the coupling strength and subsequently they become insensitive to a further increase of the coupling. This is a consequence of the rotating-wave approximation.

\section{STIRAP Photoassociation to $\Phi^{\Sigma^+_g}_{1}$}\label{second stirap}

\begin{figure}[H]
\centering
\includegraphics[width=0.6\linewidth]{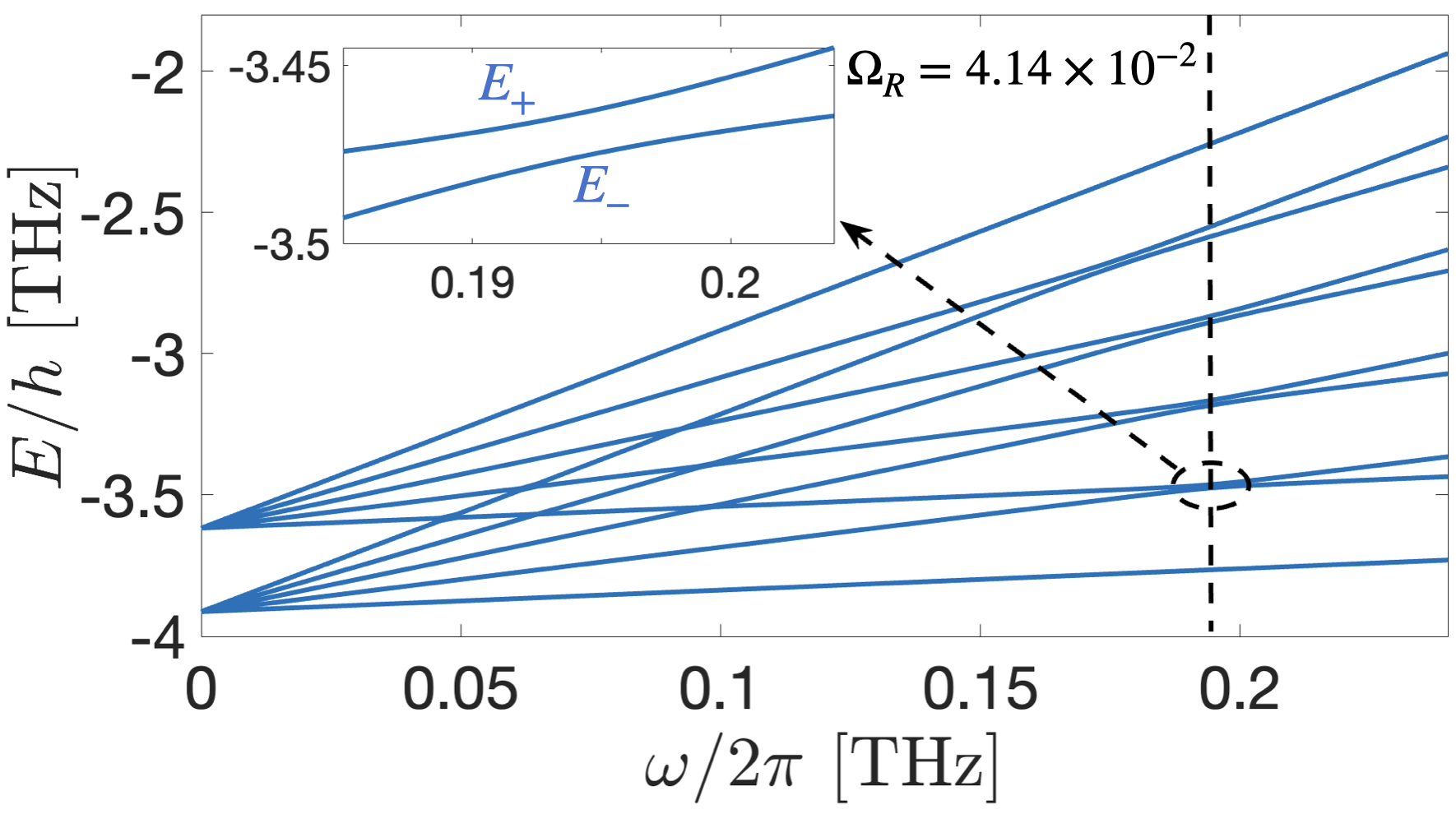}
\caption{Polaritonic energy spectra for the vibrational states $n=205$ and $m=206$ in the $\Pi_u$ PES hybridized with the vacuum cavity photons. The system is considered under strong coupling, i.e. $g=10^{-2}$, and 5 Fock states are required to converge the lowest avoided crossing shown in the inset which features a normalized Rabi splitting $\Omega_R=4.14\times 10^{-2}$. }
\label{spectra_205206}
\end{figure}

For completeness, next we provide a second example where polariton formation is advantageous for the photoasociation of $\textrm{Rb}_2$ molecules. In this case, we consider higher-lying vibrational states in the $\Pi_u$ PES, namely $\Phi^{\Pi_u}_{205}$ and $\Phi^{\Pi_u}_{206}$, aiming to photoassociate $\textrm{Rb}_2$ to the first excited vibrational state in $\Sigma^+_g$, i.e. $\Phi^{\Sigma^+_g}_{1}$. The respective energy spectrum for the vibrational states $\Phi^{\Pi_u}_{205}$ and $\Phi^{\Pi_u}_{206}$ strongly coupled to the cavity with $g=10^{-2}\textrm{a.u.}$ using the length gauge Hamiltonian is provided in Fig.~\ref{spectra_205206} taking into account five photon Fock states. A magnified version of the lowest avoided crossing featuring a normalized Rabi splitting $\Omega_R=4.14\times 10^{-2}$ which confirms the strong light-matter interaction is illustrated in the inset of  Fig.~\ref{spectra_205206}.

\begin{figure}[H]
     \centering
     \begin{subfigure}[b]{0.49\textwidth}
         \centering
        \includegraphics[width=\columnwidth]{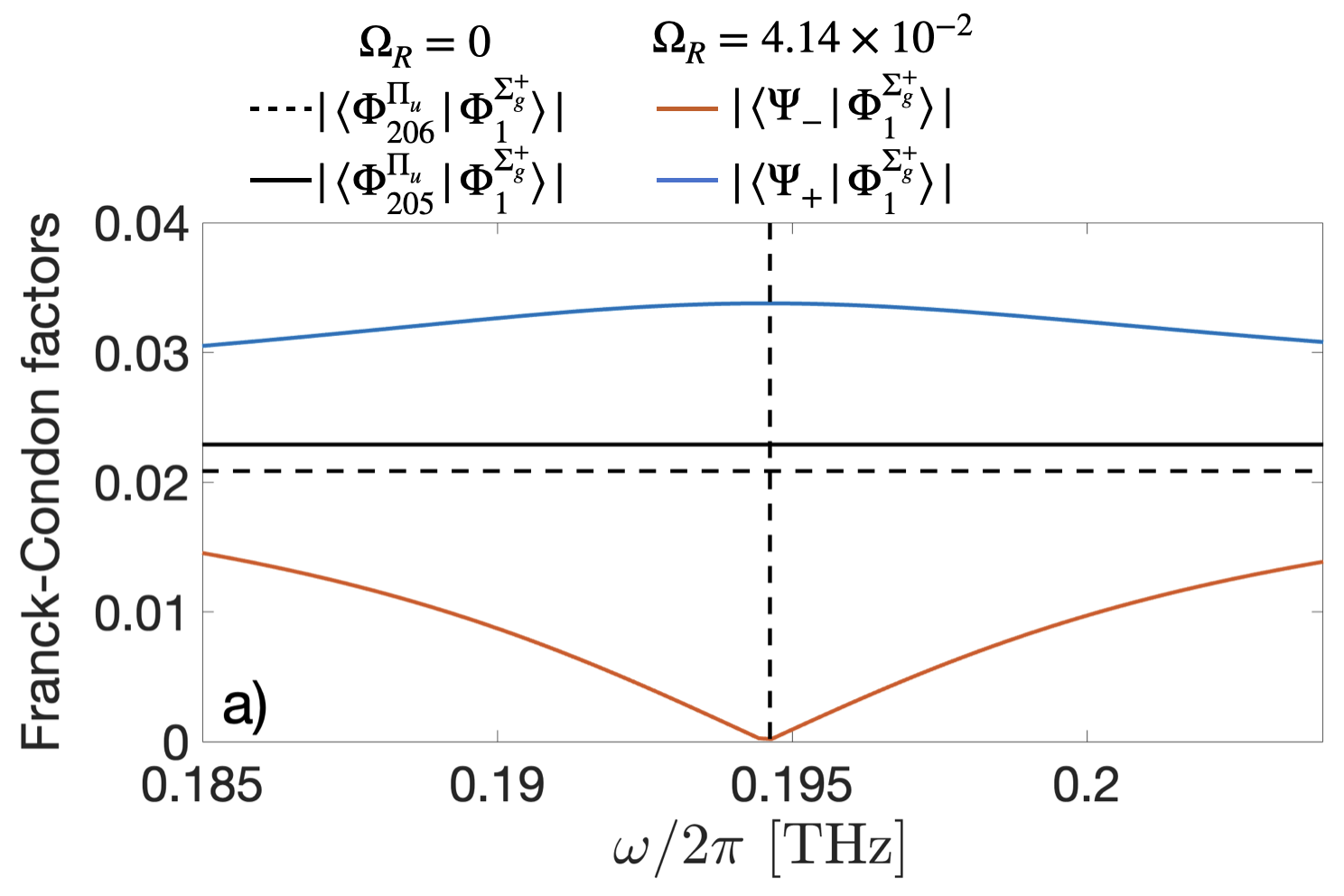}
   \end{subfigure}
     \hfill
     \begin{subfigure}[b]{0.49\textwidth}
         \centering
         \includegraphics[width=0.9\columnwidth]{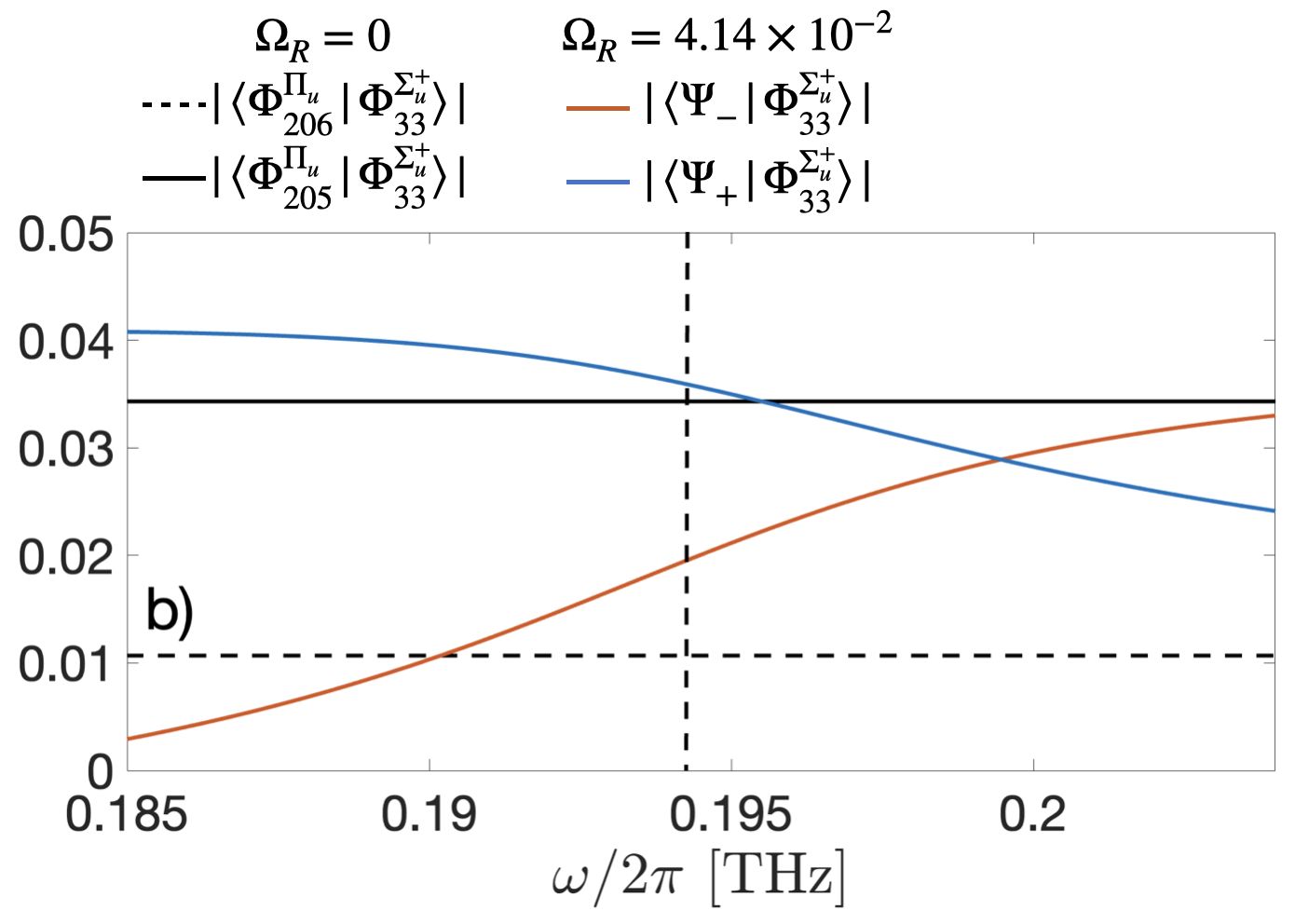}
 \end{subfigure}
\caption{Bare and polaritonic FC factors for the downwards transition a) from the states in the $\Pi_u$ PES to the $\Phi^{\Sigma^+_g}_{1}$ (see legend), and b) for the upwards transition from the state $\Phi^{\Sigma^+_u}_{33}$ to the states in the $\Pi_u$ PES (see legend). For both transitions the upper polariton exhibits enhanced FC factors around the resonance while the FC factor of the lower polariton is strongly suppressed. The system is in the strong coupling regime with $\Omega_R=4.14\times 10^{-2}$.}
\label{PFCs205}
\end{figure}

Subsequently, we compute the polaritonic FC factors for the upwards transition, namely from $\Phi^{\Sigma^+_u}_{33}$ to the states in the $\Pi_u$ PES, and the transition from the $\Pi_u$ PES to $\Phi^{\Sigma^+_g}_1$ shown in Fig.~\ref{PFCs205}. 
As it can be seen from Fig.~\ref{PFCs205}(a), throughout the whole frequency range the FC factor of the upper polariton $\Psi_+$ is significantly enhanced, while the lower polariton exhibits strong suppression, especially at resonance indicated by the vertical dashed line. Furthermore, inspecting Fig.~\ref{PFCs205}(b) it turns out that at the resonance point (vertical dashed line) and before it, the FC factor of the upper polariton is enhanced as compared to the bare state $\Phi^{\Pi_u}_{205}$, while shortly after the resonance is getting suppressed. On the other hand, the FC factor of the lower polariton $\Psi_-$ has always a lower value than the bare state $\Phi^{\Pi_u}_{205}$. Thus, it can be concluded that for both transitions the upper polariton demonstrates an enhanced FC factor at the resonance point (vertical dashed line) as compared to the bare vibrational states $\Phi^{\Pi_u}_{205,206}$. This fact will be crucial for the STIRAP photoassociation as we will argue below. 
It is important to mention that in this example it is the upper polariton that exhibits enhancement because the transition dipole matrix element $\langle \Phi^{\Pi_u}_{205}|R|\Phi^{\Pi_u}_{206}\rangle<0$ is negative.

\begin{figure}[H]
\centering
\begin{subfigure}[b]{0.49\textwidth}
\centering
\includegraphics[width=\columnwidth]{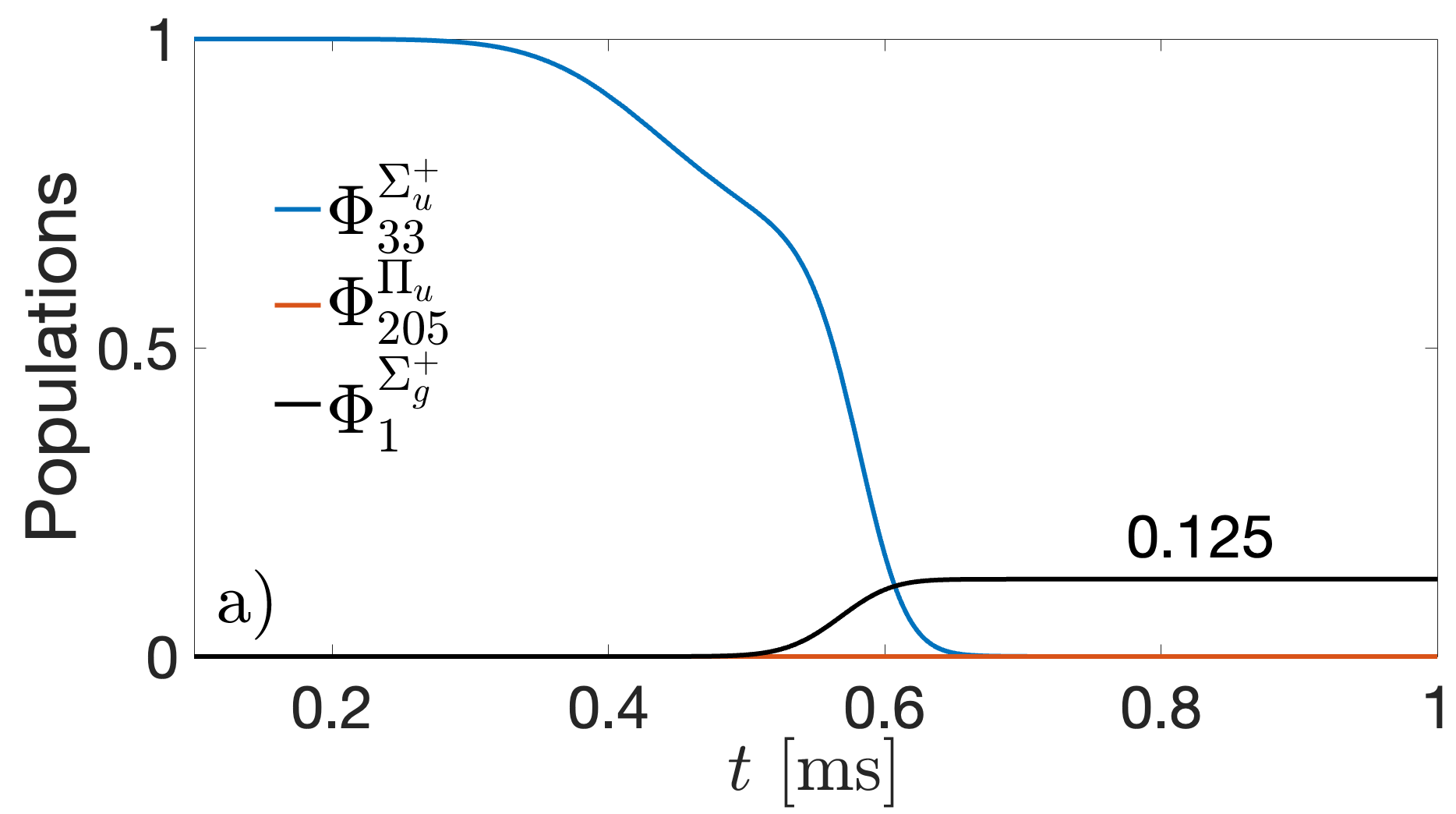}
\end{subfigure}
\hfill
\begin{subfigure}[b]{0.49\textwidth}
\centering
\includegraphics[width=\columnwidth]{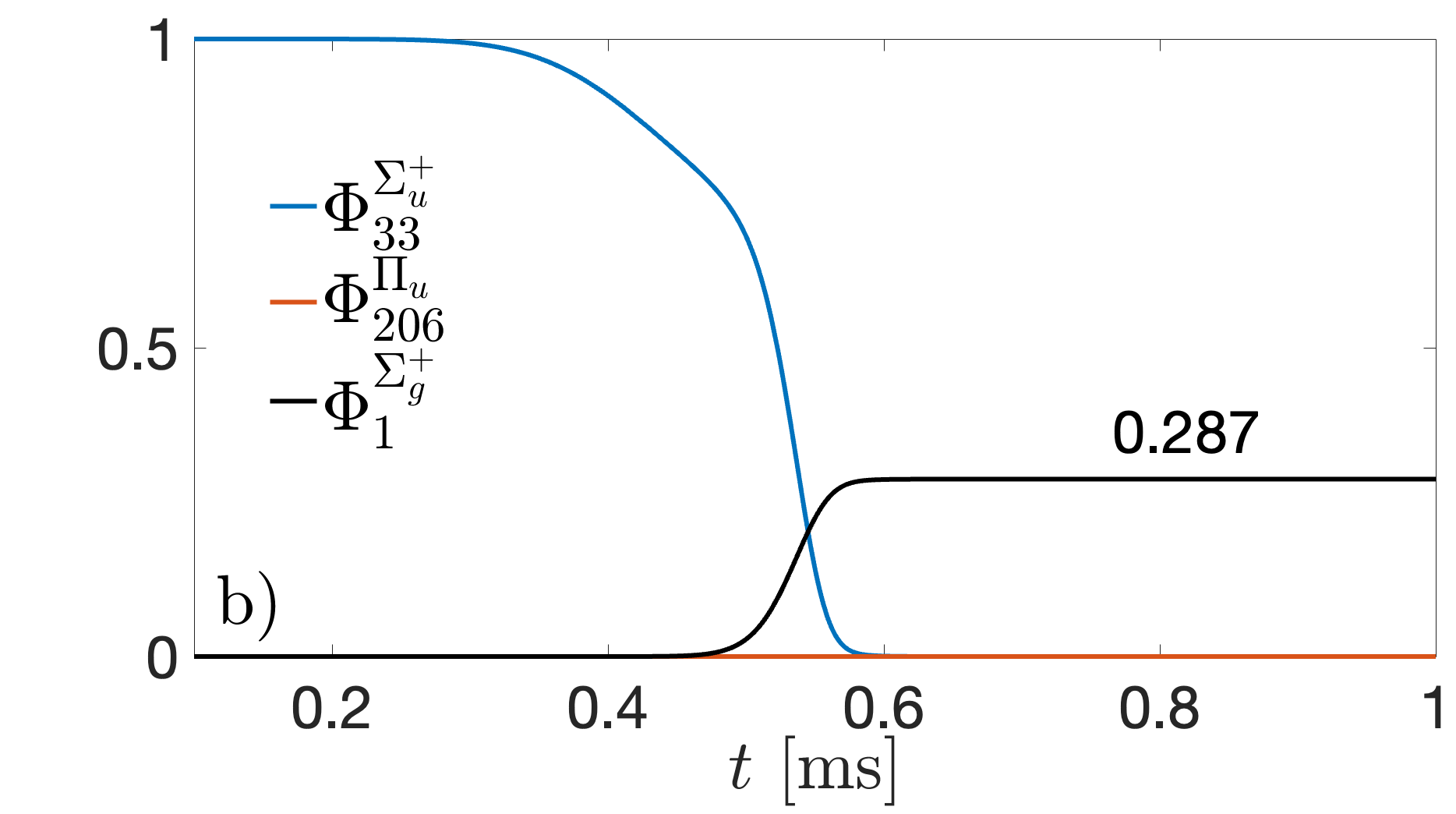}
\end{subfigure}
\caption{Temporal evolution of the populations of the three states (see legend) involved in the STIRAP. In a) the STIRAP through the state $\Phi^{\Pi_u}_{205}$ is shown and we find $12.5\%$ molecular photoassociation. In b) the STIRAP is performed via the state $\Phi^{\Pi_u}_{206}$ and $28.7\%$ molecular photoassociation is identified. }
\label{Stirap205}
\end{figure}

\begin{figure}[H]
\centering
\begin{subfigure}[b]{0.49\textwidth}
\centering
\includegraphics[width=\columnwidth]{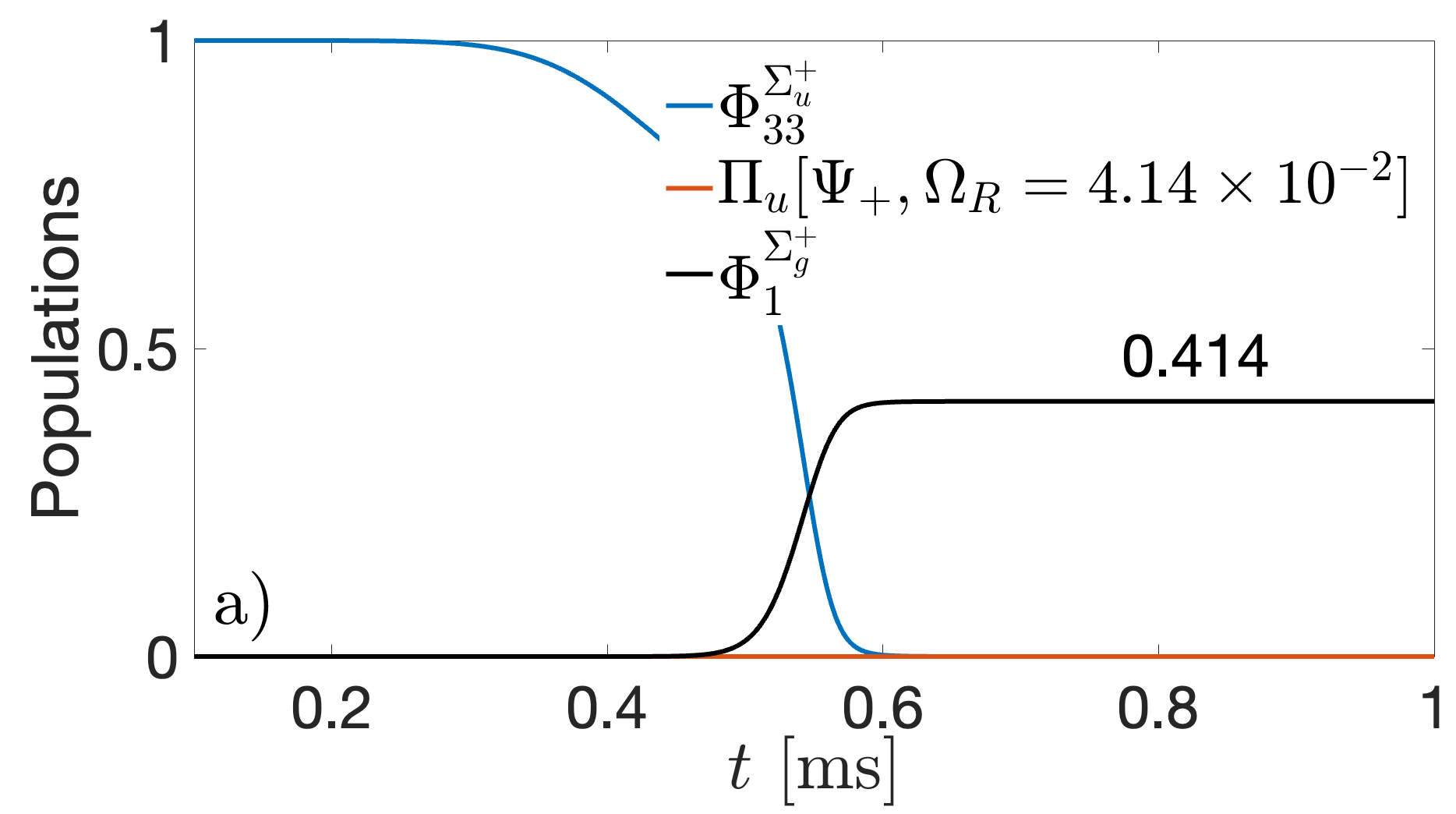}
\end{subfigure}
\hfill
\begin{subfigure}[b]{0.49\textwidth}
\centering
\includegraphics[width=\columnwidth]{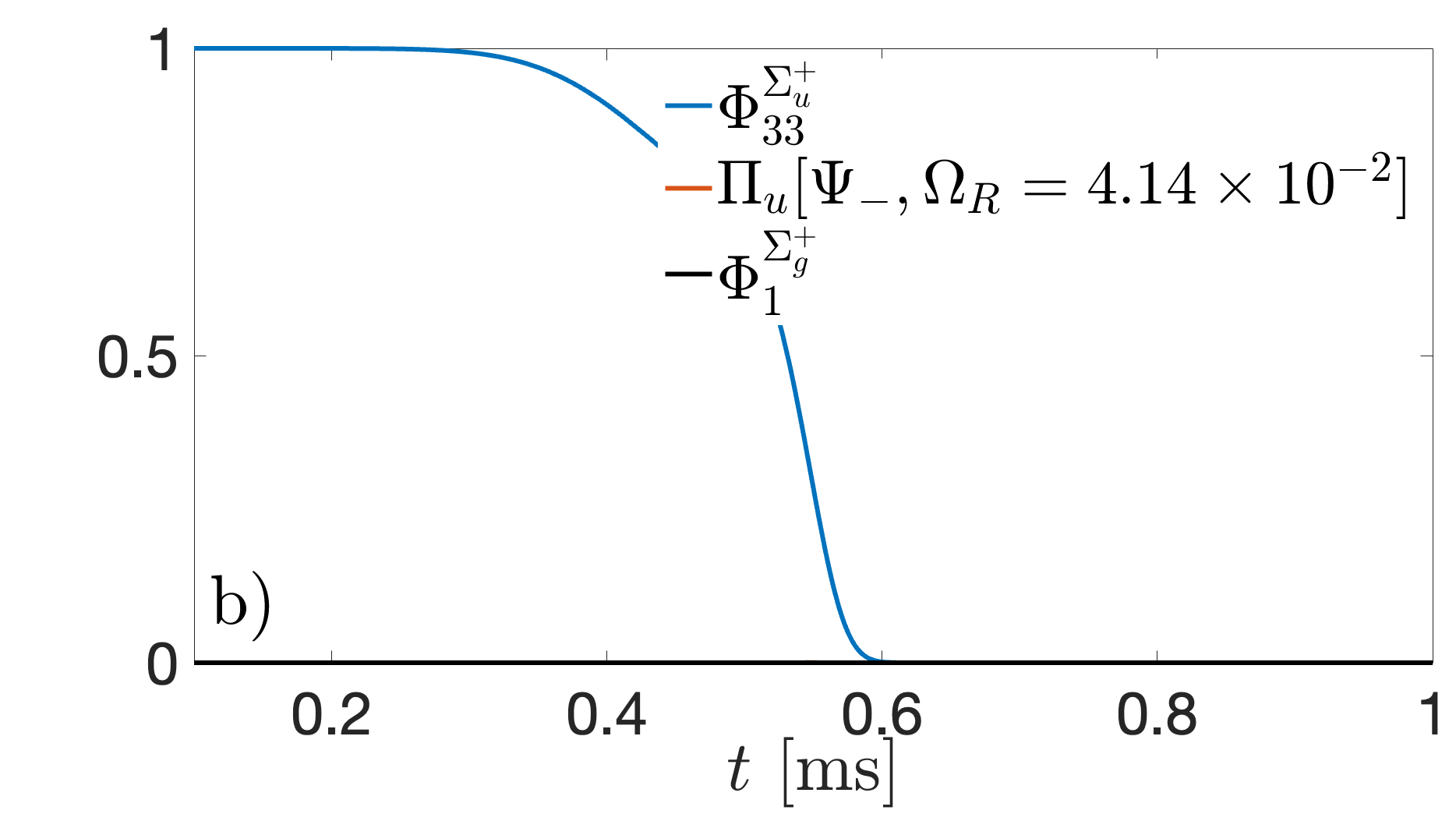}
\end{subfigure}
\caption{Time-evolution for the STIRAP photoassociation performed through the polariton states. In a) the STIRAP is performed through $\Psi_+$ resulting in $41.4\%$ molecular photoassociation, while in b) the STIRAP is made through $\Psi_-$ leading to photoassociation below $1\%$. }
\label{PolStirap205}
\end{figure}

Having obtained the polaritonic FC factors of the vibrational states $\Phi^{\Pi_u}_{205}$ and $\Phi^{\Pi_u}_{205}$ coupled to the cavity we continue by performing STIRAP utilizing the $\Lambda$-scheme described in the main text, see Eq.~(\ref{stirap}). In the $\Lambda$-scheme the state $\Phi^{\Sigma^+_u}_{33}$ is initially fully populated and it is photoassociated to $\Phi^{\Sigma^+_g}_{1}$. The molecular production using the bare vibrational states $\Phi^{\Pi_u}_{205}$ and $\Phi^{\Pi_u}_{205}$ is shown in Figs.~\ref{Stirap205} (a) and (b) where the percentage of $12.5\%$ and $28.7\%$ of molecular photoassociation respectively is observed. The corresponding populations via molecular photoassociation through the polaritonic states are presented in Fig.~\ref{PolStirap205}. It is found that the upper polariton exhibits a significantly enhanced molecular photoassociation which reaches $41.4\%$ [Fig.~\ref{PolStirap205}(a)]. On the other hand, the photoassociation through the lower polariton [Fig.~\ref{PolStirap205}(b)] is extremely suppressed and in particular much lower than $1\%$. The significantly enhanced photoassociation of the upper polariton and the respective suppression of the molecular yield in the lower polariton are traced back to the behavior of their respective FC factors shown in Fig.~\ref{PFCs205}. More precisely, in Figs.~\ref{PFCs205}(a) and (b) it was shown that at the resonance point (vertical dashed line) the upper polariton state $\Psi_+$ acquired an enhanced FC factor as compared to the bare vibrational states $\Phi^{\Pi_u}_{205}$ and $\Phi^{\Pi_u}_{205}$.

\section{Example of No Polaritonic Advantage in Photoassociation}\label{No advantage}

\begin{figure}[H]
\centering
\begin{subfigure}[b]{0.49\textwidth}
\centering
\includegraphics[width=\columnwidth]{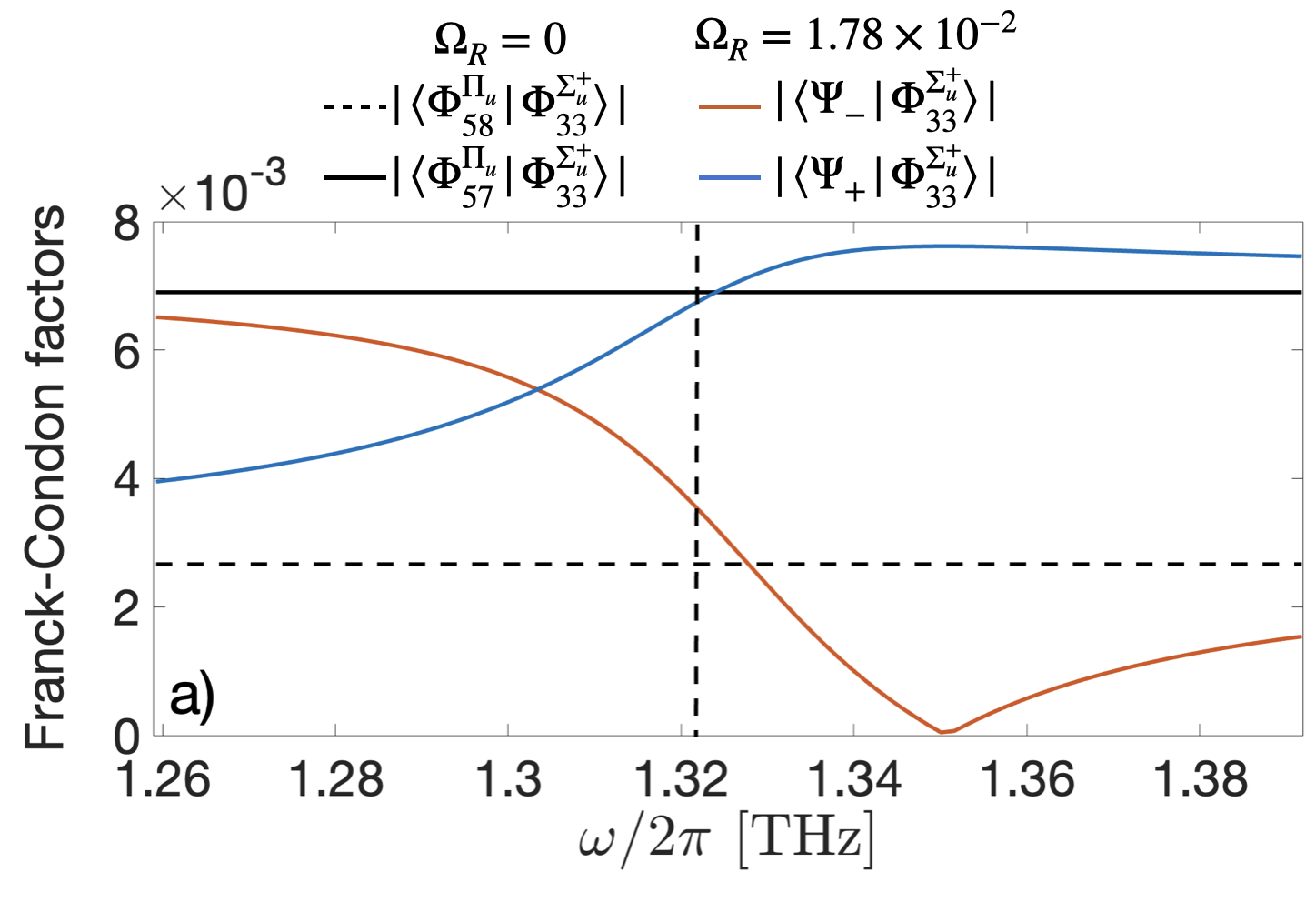}
\end{subfigure}
\hfill
\begin{subfigure}[b]{0.49\textwidth}
\centering
\includegraphics[width=\columnwidth]{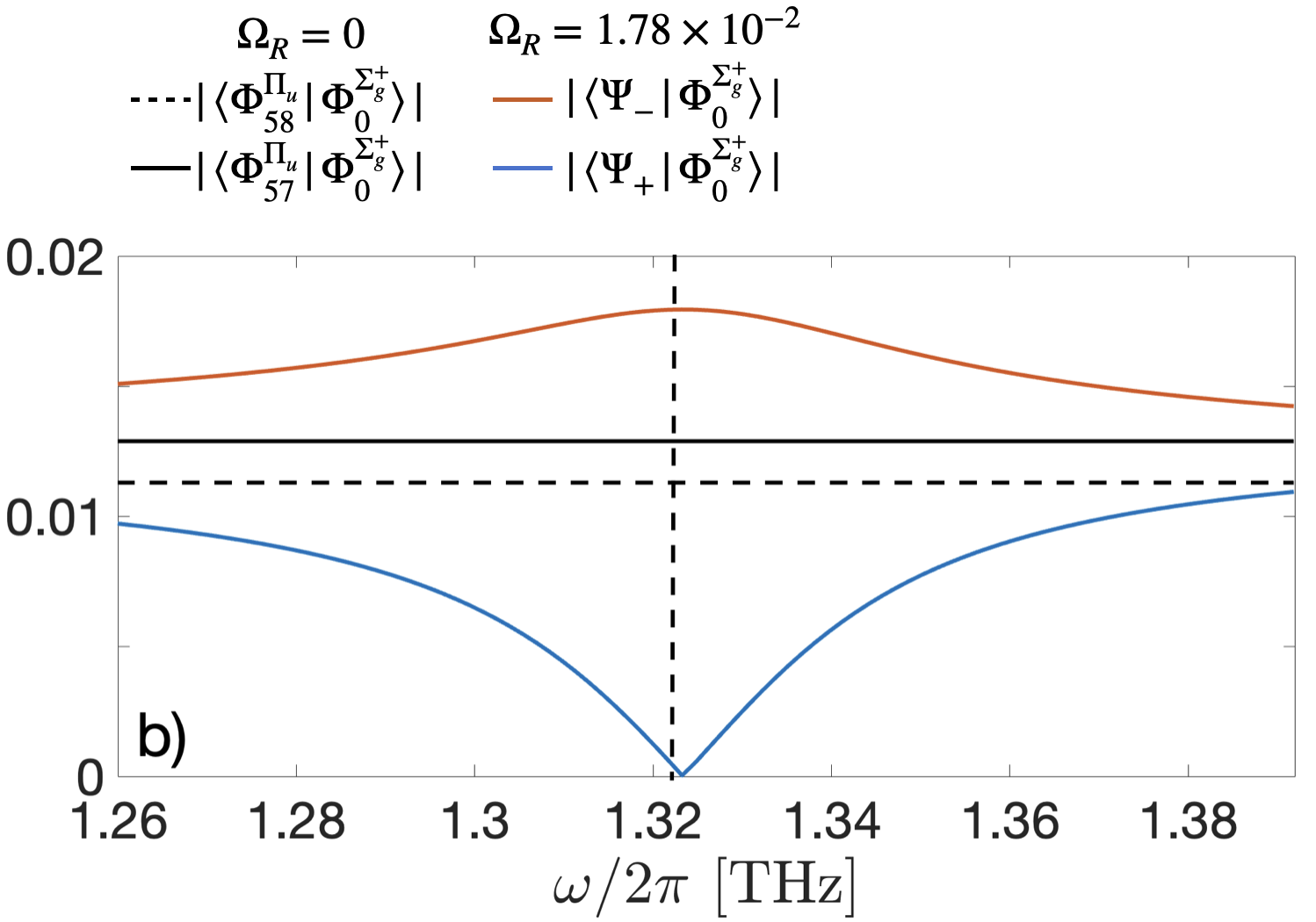}
\end{subfigure}
\caption{Bare and polaritonic FC factors for a) the upwards transition from the state $\Phi^{\Sigma^+_u}_{33}$ to the states in the $\Pi_u$ PES (see legend), and b) for the downwards transition from the states in the $\Pi_u$ PES (see legend) to the state $\Phi^{\Sigma^+_g}_{0}$. In a) the upper polariton acquires an enhanced FC factor while in b) the situation is inverted.}
\label{PFCs5758}
\end{figure}

In the main text we argued that STIRAP photoassociation through the hybrid polariton states has a clear advantage only when the FC factors of the same polariton branch are enhanced for both transitions following the $\Lambda$-scheme. Here, we provide a characteristic example where the polariton states do not exhibit an advantage for the photoassociation of molecules. We consider the molecular states $\Phi^{\Pi_u}_{57}$ and $\Phi^{\Pi_u}_{58}$ which are strongly coupled to the cavity with light-matter coupling strength $g=10^{-2}\textrm{a.u.}$. In this case, the lowest avoided crossing of the system (not shown) exhibits normalized Rabi splitting $\Omega_R=1.78\times 10^{-2}$. The resulting FC factors of the upper and lower polariton states that participate in the first avoided crossing are depicted in Fig.~\ref{PFCs5758}. It can be readily seen from  Fig.~\ref{PFCs5758}(a) that after the resonance point, indicated by the vertical dashed line, the FC factor of the upper polariton is larger than the bare ones of the molecular states, while the lower polariton has a much lower FC factor. Before the resonance point the FC factors of both polariton states are lower than the vibrational state $\Phi^{\Pi_u}_{57}$. Exactly at resonance, both polaritons have lower FC factor than the molecular state $\Phi^{\Pi_u}_{57}$. In Fig.~\ref{PFCs5758}(b) the situation is simpler and we observe that throughout the whole  frequency window where the avoided crossing occurs, the lower polariton $\Psi_-$ has a substantially larger FC factor than both bare molecular states $\Phi^{\Pi_u}_{57,58}$, while the upper polariton is significantly suppressed. At resonance in Fig.~\ref{PFCs5758}(b) we see the FC factors of the polariton states are maximally modified.

\begin{figure}[H]
\centering
\begin{subfigure}[b]{0.49\textwidth}
\centering
\includegraphics[width=\columnwidth]{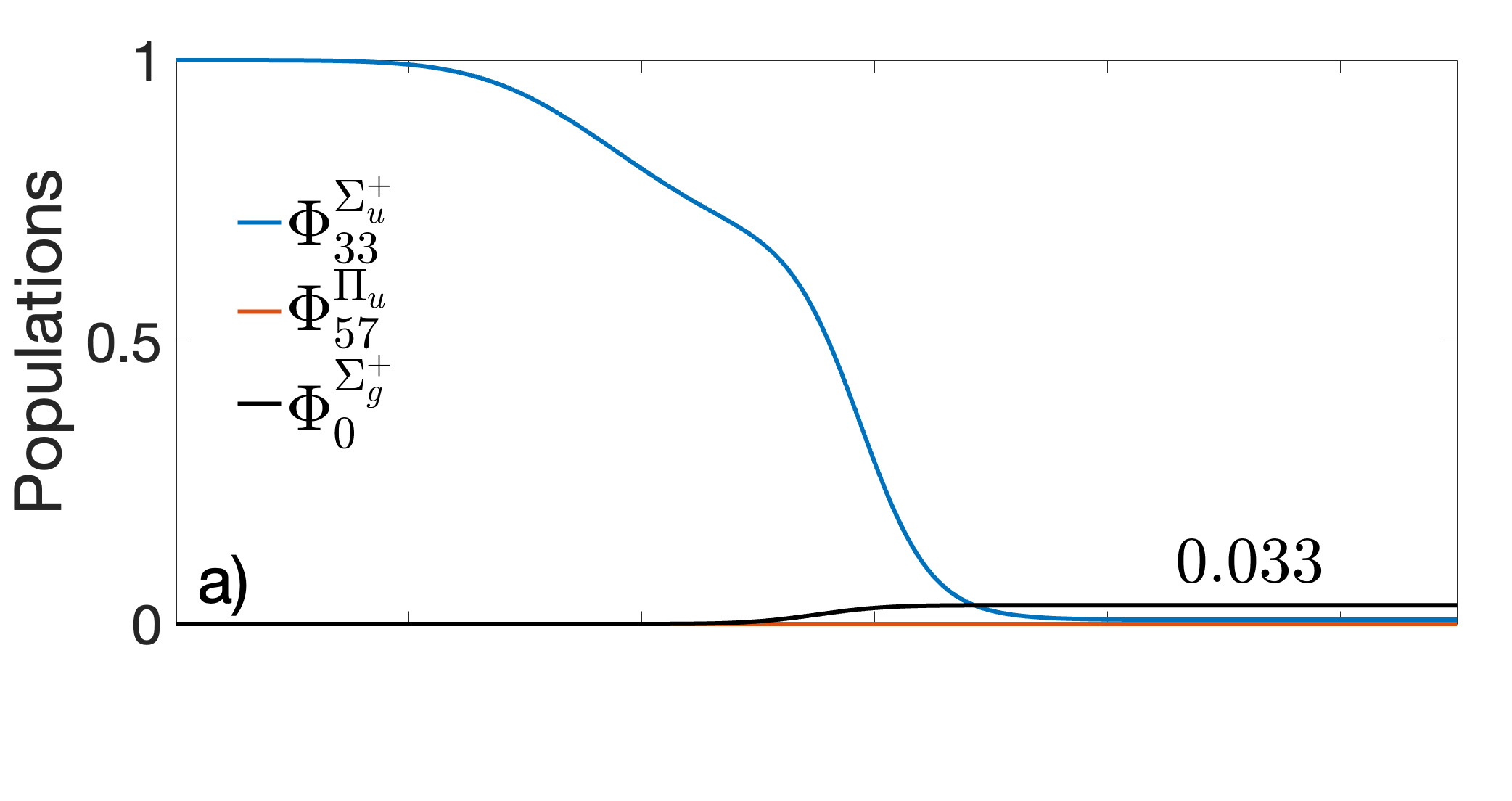}
\end{subfigure}
\hfill
\begin{subfigure}[b]{0.49\textwidth}
\centering
\includegraphics[width=\columnwidth]{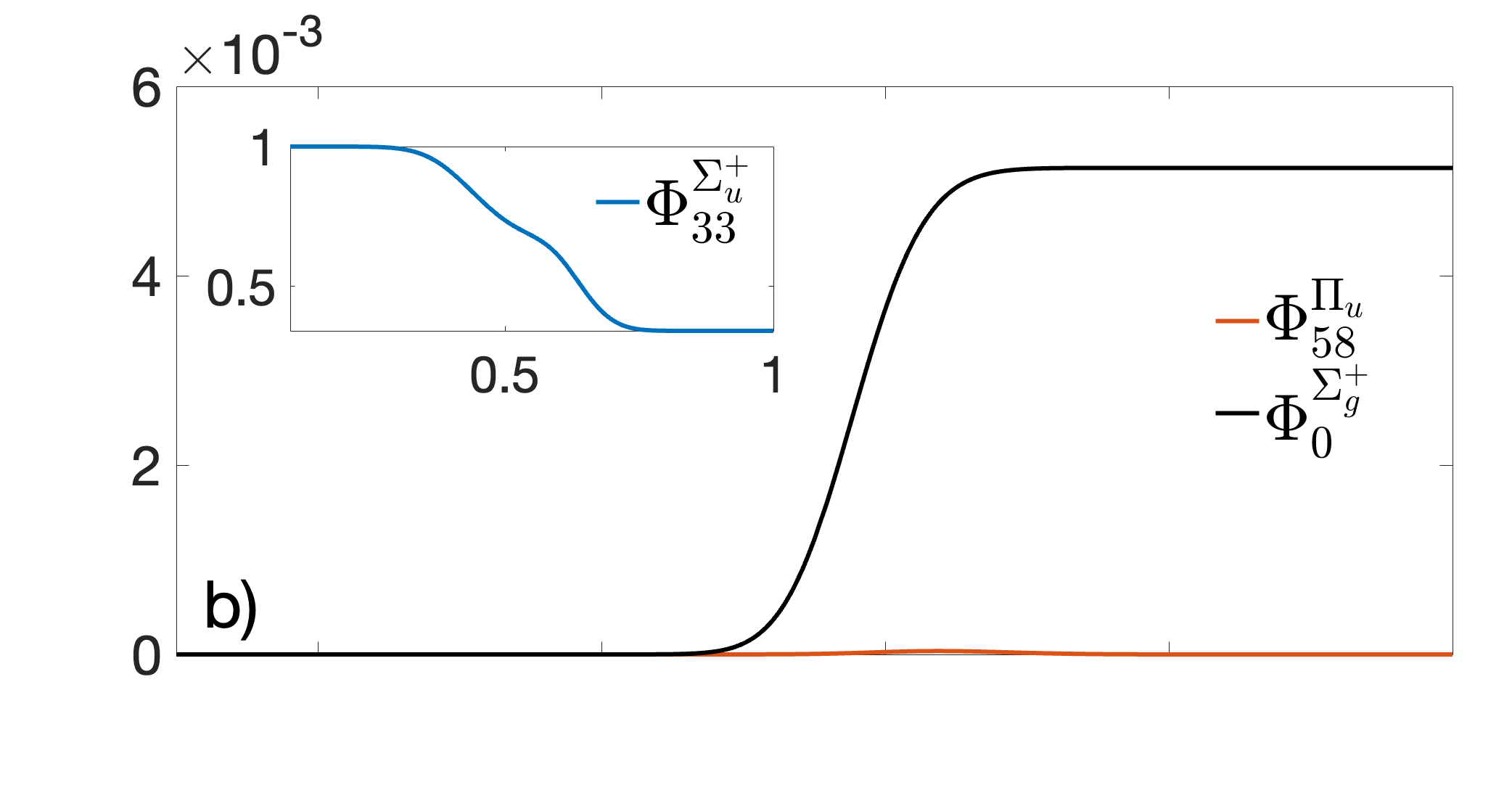}
\end{subfigure}
\hfill 
\begin{subfigure}[b]{0.49\textwidth}
\centering
\includegraphics[width=\columnwidth]{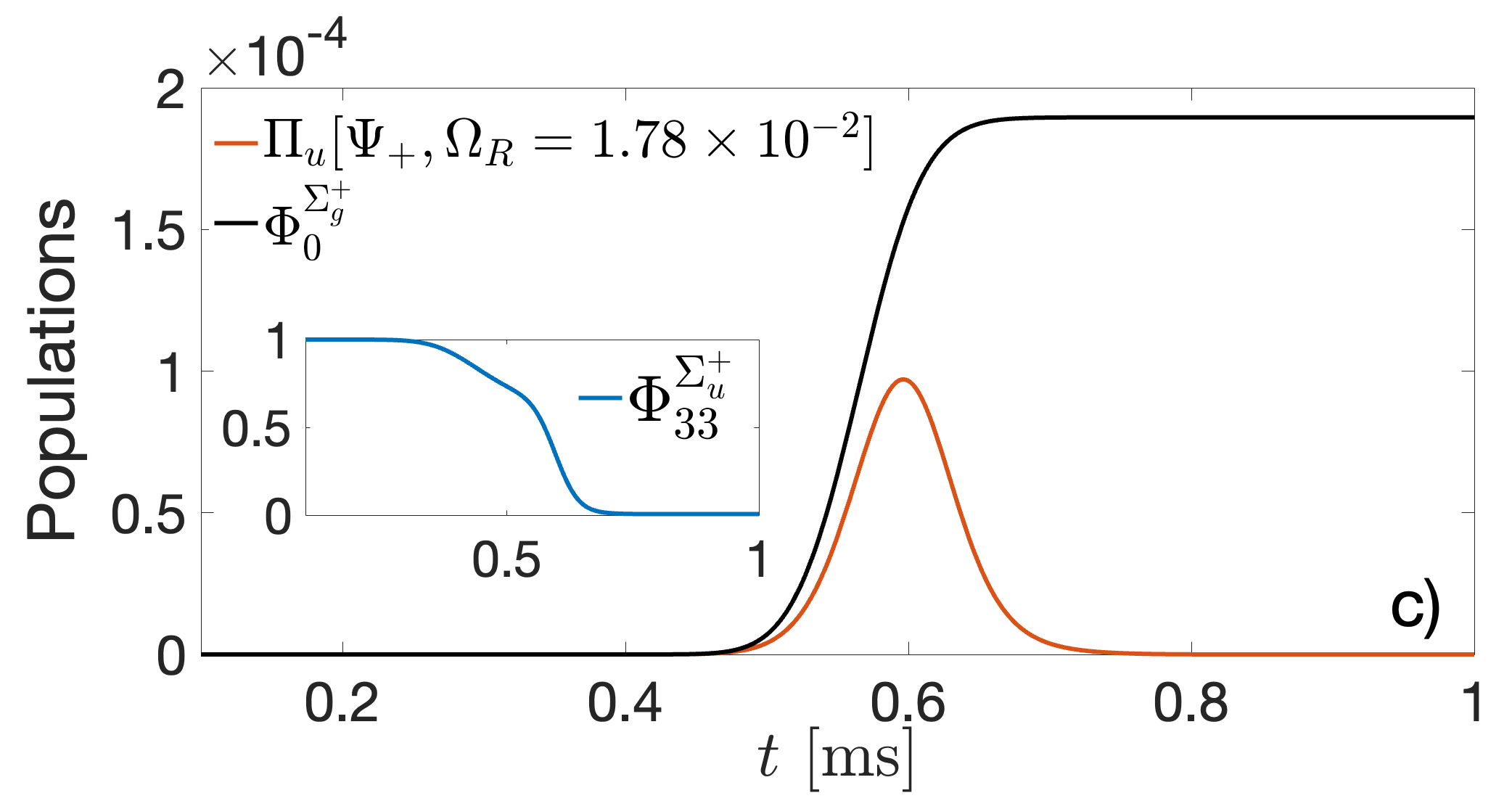}
\end{subfigure}
\hfill
\begin{subfigure}[b]{0.49\textwidth}
\centering
\includegraphics[width=\columnwidth]{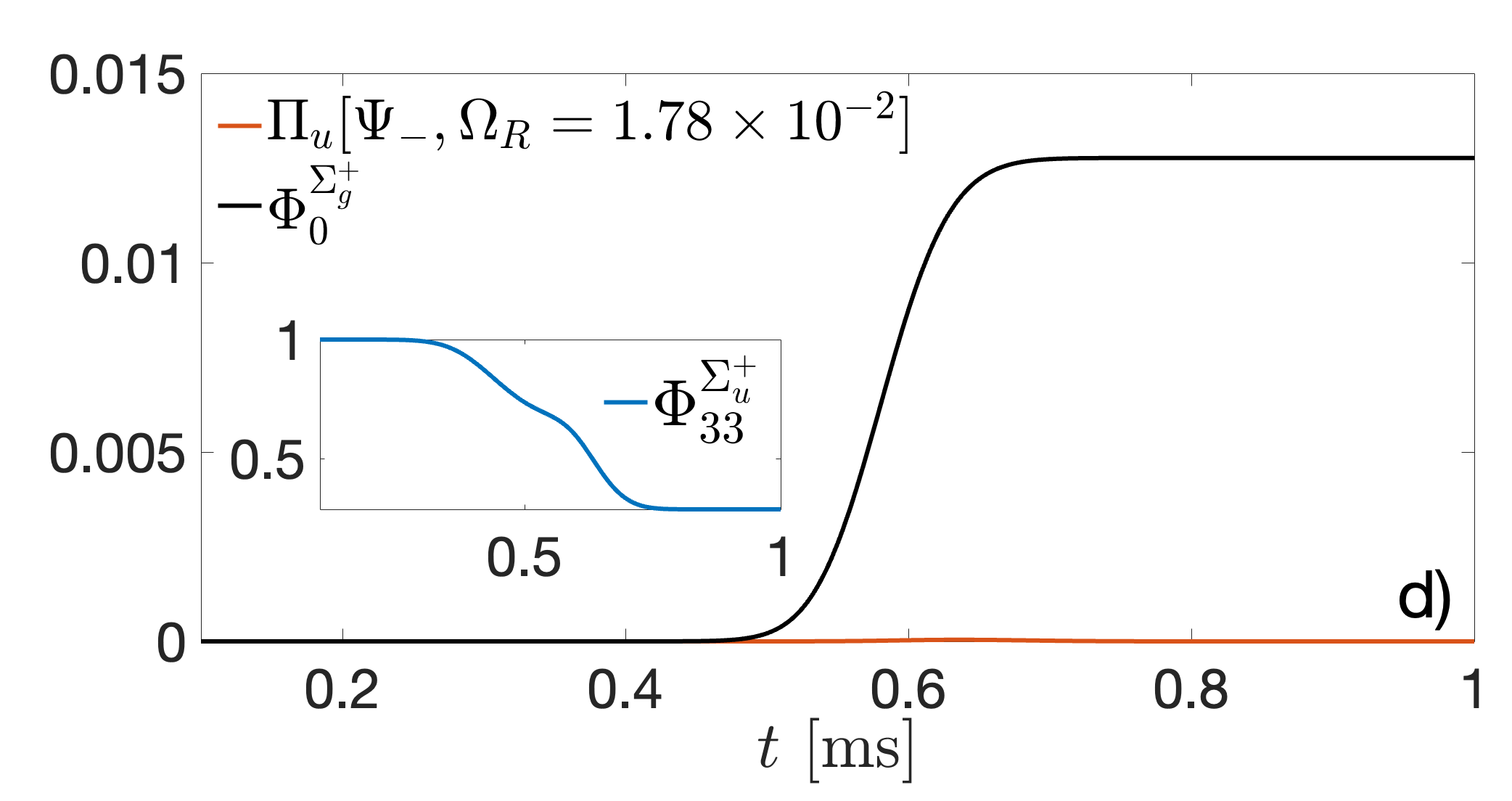}
\end{subfigure}
\caption{Time-evolution of the populations of the three states (see legends) involved in the STIRAP. In a) the STIRAP through $\Phi^{\Pi_u}_{57}$ is shown demonstrating $3.3\%$ photoassociation. In b) the STIRAP photoassociation through $\Phi^{\Pi_u}_{58}$ is $0.5\%$. In c) and d) the STIRAP for the upper ($\Psi_+$) and the lower ($\Psi_-$) polariton are shown for $\Omega_R=1.78\times 10^{-2}$ reaching $0.02\%$ and $\sim 1.5\%$ photoassociation respectively. None of the polaritons shows enhanced molecular photoassociation as compared to $\Phi^{\Pi_u}_{57}$.   }
\label{Stirap5758}
\end{figure}

Having at hand the polaritonic FC factors of the vibrational states $\Phi^{\Pi_u}_{57}$ and $\Phi^{\Pi_u}_{58}$ we continue by performing STIRAP. The corresponding $\Lambda$-scheme is initiated with the state $\Phi^{\Sigma^+_u}_{33}$ fully populated, and we photoassociate to $\Phi^{\Sigma^+_g}_{0}$. The emergent molecular production using the bare vibrational states $\Phi^{\Pi_u}_{57}$ and $\Phi^{\Pi_u}_{58}$ is shown in Figs.~\ref{Stirap5758} (a) and (b) which leads to $3.3\%$ and $\sim 0.5\%$ of molecular photoassociation respectively. Importantly, by considering the STIRAP photoassociation through the polaritonic states we observe $\sim 0.02\%$ molecular yield for the upper polariton [Fig.~\ref{Stirap5758}(c)] and  $\sim 1.5\%$ for the lower polariton [Fig.~\ref{Stirap5758}(d)]. Thus, neither of the polariton states shows an advantage for the photoassociation of $\textrm{Rb}_2$ molecules. This is the case because none of the polariton branches exhibits enhanced FC factors for both transitions in the $\Lambda$-scheme.

\end{document}